\documentclass[nofootinbib]{revtex4}
\usepackage[T1]{fontenc}
\usepackage{amsmath,amssymb}
\usepackage{epsfig}
\usepackage{dcolumn}
\usepackage{graphicx}
\usepackage[usenames,dvipsnames]{color}
\usepackage{slashed}
\usepackage[colorlinks,citecolor=blue]{hyperref}
\usepackage{pdfpages}
\usepackage{float}
\usepackage{adjustbox}
\usepackage[autostyle]{csquotes}
\begin{document}
\title{\boldmath Impact of Nuclear effects in Energy Reconstruction Methods on Sensitivity of Neutrino Oscillation Parameters at NO$\nu$A experiment}
\author{ Paramita Deka$^{1}$ \footnote{E-mail: deka.paramita@gmail.com}, Jaydip Singh$^{2}$ \footnote{E-mail: jdsingh@fnal.gov}, Kalpana Bora$^{1}$ \footnote{kalpana@gauhati.ac.in}} 

\affiliation{Department Of Physics, Gauhati University, Assam, India$^{1}$}
\affiliation{Department Of Physics, Lucknow University, Uttar Pradesh, India$^{2}$}

\begin{abstract}

Long baseline (LBL) neutrino experiments aim to measure the neutrino oscillation parameters to high precision. These experiments use nuclear targets for neutrino scattering and hence are inflicted with complexities of nuclear effects. Nuclear effects and their percolation into sensitivity measurement of neutrino oscillations parameters are not yet fully understood and therefore need to be dealt with carefully. In a recent work \cite{Deka:2021qnw}, we reported some results on this for NO$\nu$A experiment using the kinematic method of neutrino energy reconstruction, where it was observed that the nuclear effects are important in sensitivity analysis, and inclusion of realistic detector set up specifications increases uncertainty in this analysis as compared to ideal detector case. Precise reconstruction of neutrino energy is an important component in the extraction of oscillation parameters, and it is required that the neutrino energy is reconstructed with very high precision. With this motivation, in this work, we use two methods of neutrino energy reconstruction - kinematic and calorimetric, including the nuclear effects, and study their impact on sensitivity analysis. We consider nuclear interactions such as RPA and 2p2h and compare two energy reconstruction methods with reference to events generation, measurement of neutrino oscillation parameters $\Delta m_{32}^2$ and $\theta_{23}$ for ($\nu_{\mu}\rightarrow\nu_{\mu}$) disappearance channel, mass hierarchy sensitivity, and CP-violation sensitivity for ($\nu_{\mu}\rightarrow\nu_{e}$) appearance channel of the NO$\nu$A experiment. It is observed that with an ideal detector setup, the kinematic method shows significant dependence on nuclear effects, as compared to the calorimetric method. We also investigate the impact of realistic detector setup for NO$\nu$A in these two methods (with nuclear effects) and find that the calorimetric method shows more bias (uncertainty increases) in sensitivity contours, as compared to the kinematic method. This is found to be true for both the mass hierarchies and for both neutrino and antineutrino incoming beams. The results of this study can offer useful insights into future neutrino oscillation analysis at LBL experiments (including leptonic CP phase measurement), especially when we are in the precision measurement era for the same, particularly in the context of some nuclear effects and energy reconstruction methods. 

\end{abstract}
\maketitle

\section{\label{sec:level1}Introduction}

In LBL neutrino oscillation experiments \cite{Fukuda:1998mi, Fukuda:2002pe, Ahmad:2002jz, Eguchi:2002dm, Michael:2006rx, Abe:2011fz, An:2012eh, Ahn:2012nd, Abe:2017vif, DUNE:2020txw, Ayres:2004js}, the precise measurement of neutrino energy is one of the essential ingredients to extract the neutrino oscillation parameters. As the neutrino oscillation parameters such as mixing angles, mass splittings, and the CP-violating (CPV) Dirac phase are determined from the collected energy-dependent event distribution, their precision measurement requires an accurate reconstruction of neutrino energy from the measured kinematics of the final state particles. The next generation oscillation experiments are aiming to measure the mass splittings with a precision exceeding 1\%, and the CPV phase as well, for which they need to measure the neutrino energy with a precision exceeding 1\% \cite{Ankowski:2017uvv, DUNE:2015lol}. Depending on the length of the baseline, the relevant neutrino energies have a range from a few hundred MeV to a few GeV, at which the dominant interaction processes change from Quasi-elastic (QE) to resonant and non-resonant meson production \cite{Formaggio:2013kya}. As neutrinos are produced in a secondary decay of primarily produced hadrons, their beams are quite broad in energy. Therefore the incoming neutrino energy can not be measured a priori in ongoing experiments. The reconstructed energy of the neutrino is determined on an event-by-event basis. These experiments depend on two methods for energy reconstruction - kinematic and calorimetric method \cite{Ankowski:2015jya}. The calorimetric method plays an important role at higher energies, which is estimated by adding the energies of all the observed particles in the detector while the kinematic method has been used to determine the incoming neutrino energy assuming quasi-free kinematics of outgoing lepton in the QE process at lower energies (a few hundred MeV to a few GeV). \\

Modern neutrino experiments do not use free nucleons as the primary target, instead, they use heavy nuclear targets (like carbon, oxygen, argon, etc.), in which complexities regarding nuclear effects are unavoidable. Nuclear effects play an important role and contribute significantly to the incorrect estimation of neutrino energy as they lead to a significant amount of missing energy during Final State Interactions (FSIs). Among all the uncertainties, nuclear effects are considered one of the largest sources of systematic uncertainties in the oscillation analysis of LBL experiments. Nuclear effects can be broadly divided into two categories - initial-state and final-state effects. Initial-state effects affect the nucleon before the neutrino interactions while hadrons produced by final-state effects influence the outgoing final-state particles before their exit from the nucleus. A true charged current (CC) QE process is represented as $\nu n\rightarrow\ l^{-} p$, where $l$ is a charged lepton. These true interaction processes are accompanied by some other processes where the outgoing proton re-interacts inside the nucleus thus producing $\Delta$ resonance. This $\Delta$ then decays to produce a pion which is then absorbed in the nucleus through FSIs. This can be represented as: $\nu p\rightarrow\mu^{-}p\pi^{+}$ or $\nu p\rightarrow\mu^{-}\Delta^{++}$. Thus the absence of the pion in the final state (called "stuck pion") leads to missing energy and appears as a QE-like event. Therefore because of FSI, a non-QE even may be wrongly identified as QE. A second complication arises due to the presence of multi-nucleon events in which the incoming neutrino interacts with, e.g, two nucleons (so-called 2p2h) \cite{Martini:2011wp, Martini:2012uc, Martini:2012fa, Benhar:2013bwa, Lalakulich:2012ac}. The neutrino energy reconstructed in such events differs significantly from their true energy value. Though pion production contributes to the background in any QE process, later it has been shown that 2p2h excitations and some other processes also shift the reconstructed energy towards lower energy bins. In MiniBooNE and K2K experiments, it was found that QE contains about 30\% contribution from 2p2h events \cite{Martini:2009uj, Martini:2011wp, Martini:2010ex}. The effect of these uncertainties in reconstructed energy on extracting the neutrino oscillation parameters for MiniBooNE and T2K experiments can be found in Refs. \cite{Lalakulich:2012hs, Meloni:2012fq, Coloma:2013rqa}.\\

\begin{figure}[H]
 \centering\includegraphics[width=9.5cm, height=6cm]{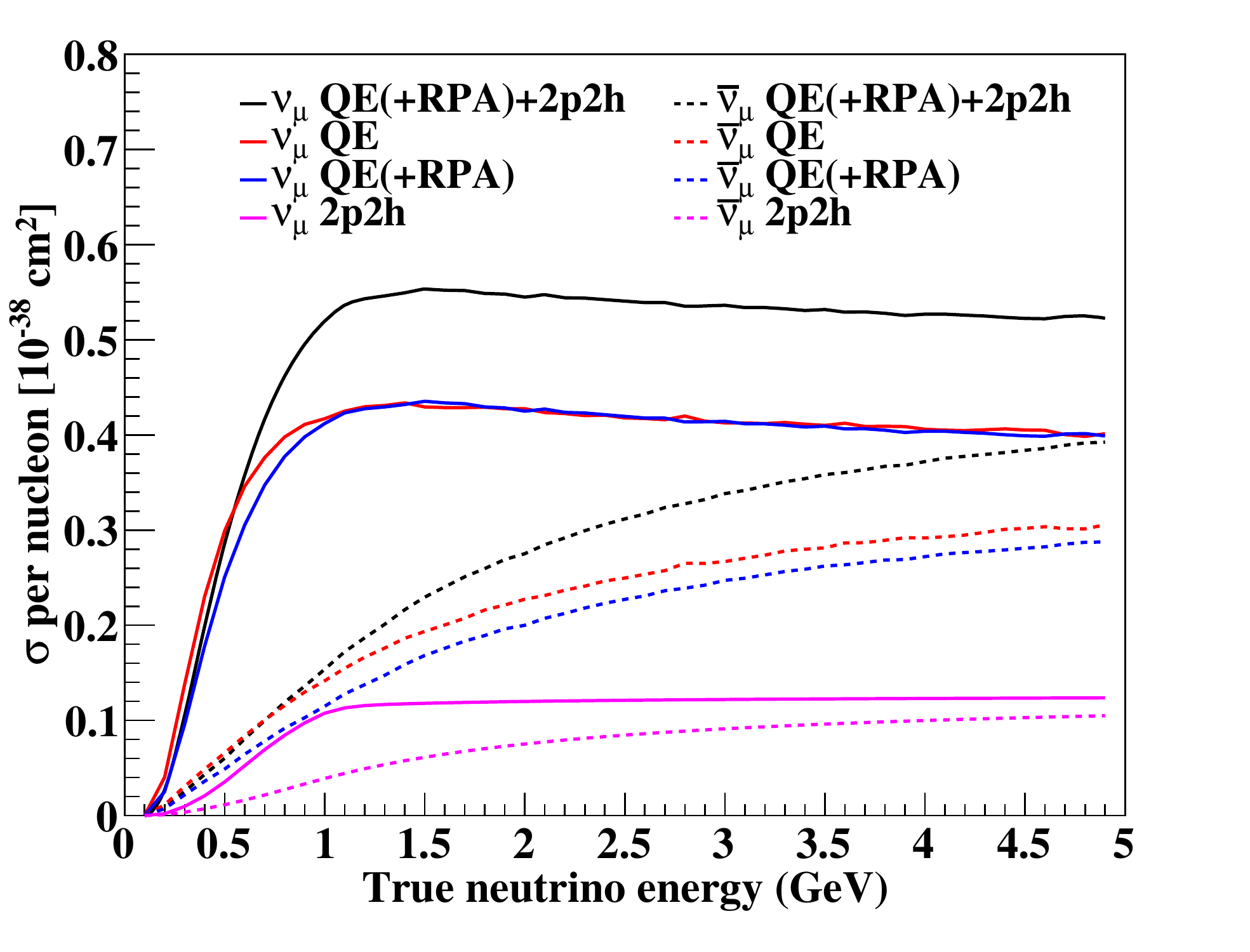}
 \caption{Cross-section as a function of true neutrino energy for both neutrino (solid line) and antineutrino (dashed line). The color scheme for different interactions: QE(+RPA)+2p2h (black line), QE only (red line), QE(+RPA) (blue line), and 2p2h only (magenta line).}
\end{figure}

We would like to mention that in \cite{Ankowski:2017uvv}, the relevance of nuclear effects (FSI) in the calorimetric reconstruction of neutrino energy was studied and found that when detector effects are included, bias on sensitivity analysis is more, than the case when detector effects are not included/ideal detector set-up. In \cite{ Ankowski:2015jya}, they did a comparative study of the calorimetric and kinematic methods of neutrino energy reconstruction in disappearance experiments, and the chosen detector capabilities were not to represent any existing detector but were indicative of the general level of performance. They found that the calorimetric method is less sensitive to the nuclear model, but is strongly affected by detector effects. While, the kinematic method depends strongly on the nuclear model, and is less sensitive to detector effects. \\

In this work, we consider realistic detector details of NO$\nu$A experiment, and use detector resolution of NO$\nu$A and two values of detector efficiency - 31.2\% (33.9\%) for $\nu_{\mu}$($\bar\nu_{\mu}$) (realistic value of NO$\nu$A ) and 100\%, and compare them. Our nuclear interaction models also differ from those of \cite{ Ankowski:2015jya}. We consider the RPA suppression and the multi-nucleon enhancement at the same time to obtain a complete model QE(+RPA)+2p2h to describe the QE interactions for both neutrino and antineutrino for both the mass hierarchies, which are believed to be more appropriate and complete models for realistic neutrino-nucleus scattering \cite{Deka:2021qnw}. In \cite{ Ankowski:2015jya} they used the Relativistic Fermi Gas (RFG) model and spectral function (SF) approach, while we have used the local Fermi gas model. We have used the Valencia model (Nieves et. al. \cite{Nieves:2012yz}), while in \cite{ Ankowski:2015jya} Dytman model \cite{Katori:2013eoa} was used. We have shown migration matrices for both reconstruction methods, for the four interaction processes QE(+RPA)+2p2h, QE, QE(+RPA), and 2p2h to have a clear visual impression of the smearing effects. Feldman-Cousins approach \cite{Feldman:1997qc} is used to compute the systematic uncertainties in sensitivity of neutrino oscillation parameter measurement, calculate $\chi^{2}$ and determine the confidence regions in ($\theta^2_{23}-\Delta m^2_{32}$) plane. Hence, though qualitatively our results do not refute those of \cite{ Ankowski:2015jya}, the results presented here add additional and new information to those presented in \cite{ Ankowski:2015jya}, with specific reference to NO$\nu$A, as we have utilized the latest nuclear models and realistic detector specifications of NO$\nu$A. This work is an extension of our previous work \cite{Deka:2021qnw} in which we investigated the uncertainties in the neutrino oscillation parameters measurement due to multi-nucleon effects at NO$\nu$A experiment, where we used only the kinematic method for the reconstruction of neutrino energy. Since reconstructed neutrino energy may change with the reconstruction method, therefore, in this work, we have undertaken to use and compare kinematic and calorimetric methods to reconstruct the neutrino energy, including nuclear effects such as RPA and 2p2h interactions. We compare near detector (ND) and far detector (FD) events in the two cases, and then study the sensitivity of these methods to nuclear effects in the $\nu_{\mu}$ disappearance channel. We compute and show the impact of 2p2h processes, RPA corrections separately in cross-section, events, and sensitivity in oscillation parameter measurement along with the QE process, using both the reconstruction methods.\\

\begin{figure}[H]
\centering\includegraphics[width=13.5cm, height=8cm]{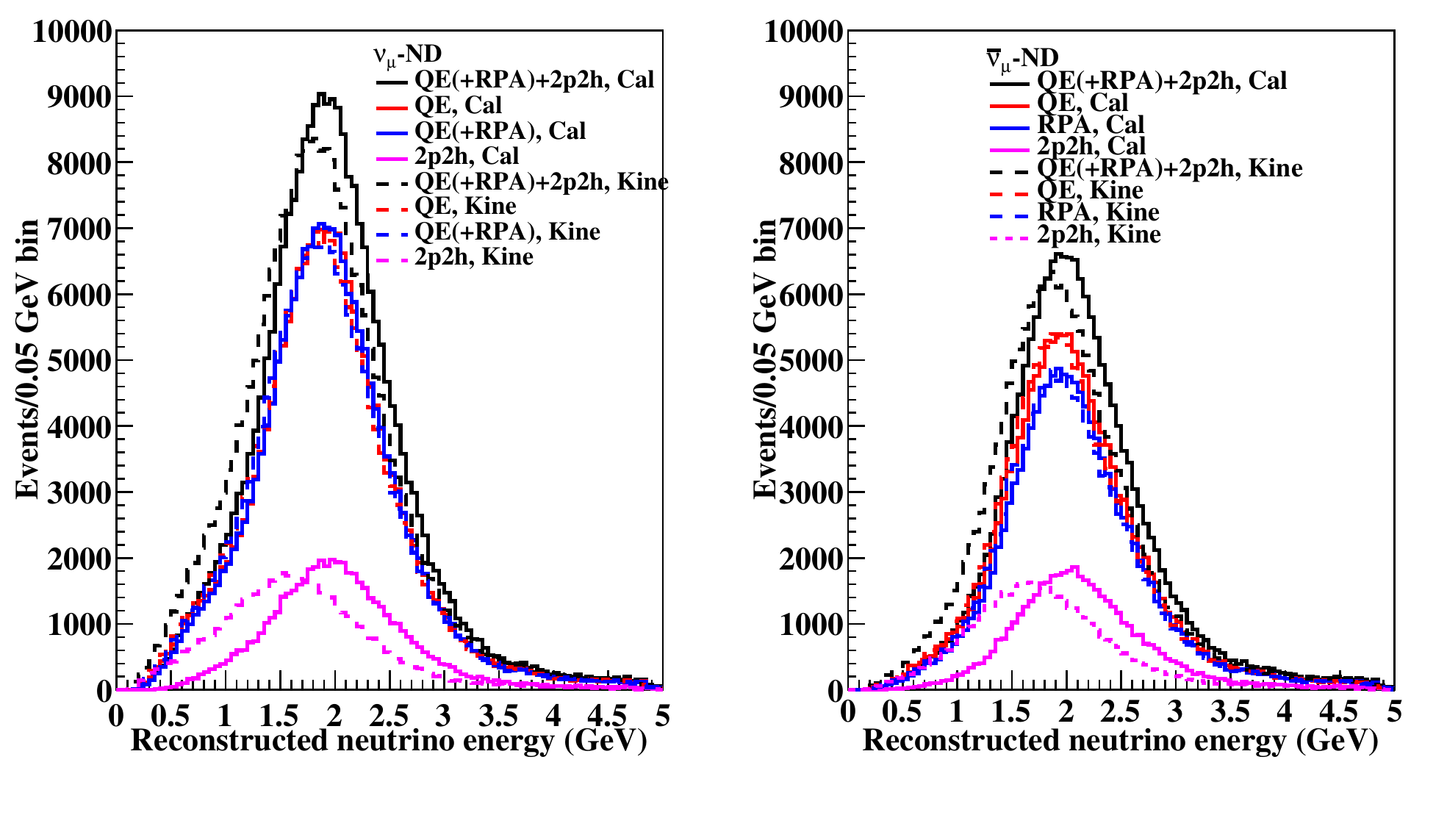}
\caption{Comparison of ND events as a function of the calorimetric (solid line) and kinematic (dashed line) reconstruction energy for neutrino (left panel) and antineutrino (right panel) for different interactions: QE(+RPA)+2p2h (black line), QE only (red line), QE(+RPA) (blue line) and 2p2h only (magenta line).}
\end{figure} 

The paper is organized as follows. In section II, we briefly review the NO$\nu$A experiment and in section III, the principle of the two energy reconstruction methods is outlined. The CP-violation sensitivity and the determination of mass hierarchy are discussed in section IV. This is followed by a description of simulation details of the neutrino event generator GENIE in section V. In section VI, we perform a detailed comparison of both the reconstruction methods in various interaction processes including the multi-nucleon effects and present sensitivity contours in three regions of parameter spaces - ($\theta^2_{23}-\Delta m^2_{32}$) plane, CP-violation, and mass hierarchy determination. The paper is summarised in section VII.

\section{The NO$\nu$A Experiment}
\label{sec:1}

Here, some details about the NO$\nu$A (NuMI Off-Axis $\nu_{e}$ Appearance) \cite{NOvA:2016vij} experiment are presented for the sake of completeness of this work. It is an LBL neutrino oscillation experiment which is designed to measure $\nu_{\mu}$ disappearance P($\nu_{\mu}(\bar\nu_{\mu})\rightarrow\nu_{\mu}(\bar\nu_{\mu}$)) and $\nu_{e}$ appearance P($\nu_{\mu}(\bar\nu_{\mu})\rightarrow\nu_{e}(\bar\nu_{e})$) probability with two functionally equivalent and identical detectors - 290 ton Near Detector and 14 kton Far Detector. Both detectors are situated off the central beam axis and FD is placed 14.6 mrad off-axis from the central beam so that the narrow neutrino energy flux peaks at around 2 GeV, near the first oscillation maximum driven by $\Delta m^{2}_{32}=2.5\times10^{-3} eV^{2}$. This off-axis configuration enhances the $\nu_{e}$ oscillation probability in the appearance channel. The NuMI \cite{Adamson:2015dkw} (Neutrinos at the Main Injector) beam produces neutrinos at the Fermi National Accelerator Laboratory which is measured by the ND placed at a distance of 1 km from the source and 105 m underground. NO$\nu$A measures oscillations by comparing the un-oscillated energy spectra of ND and oscillated spectra of FD located near Ash River, Minnesota at a distance of 810 km from the production target. NuMI beam is produced by using 120 GeV protons from Main Injector impinging on a fixed graphite target producing pions and kaons. Using magnetic horns, pions and kaons of desired charge and momentum are focused into a narrow beam where these particles further spontaneously decay to the muons, anti-muons, and their neutrino partners. Muons and anti-muons are filtered out from the decay pipe with 240 meters thick rock wall. Both the 

\begin{figure}[H] 
\centering\includegraphics[width=13.5cm, height=8cm]{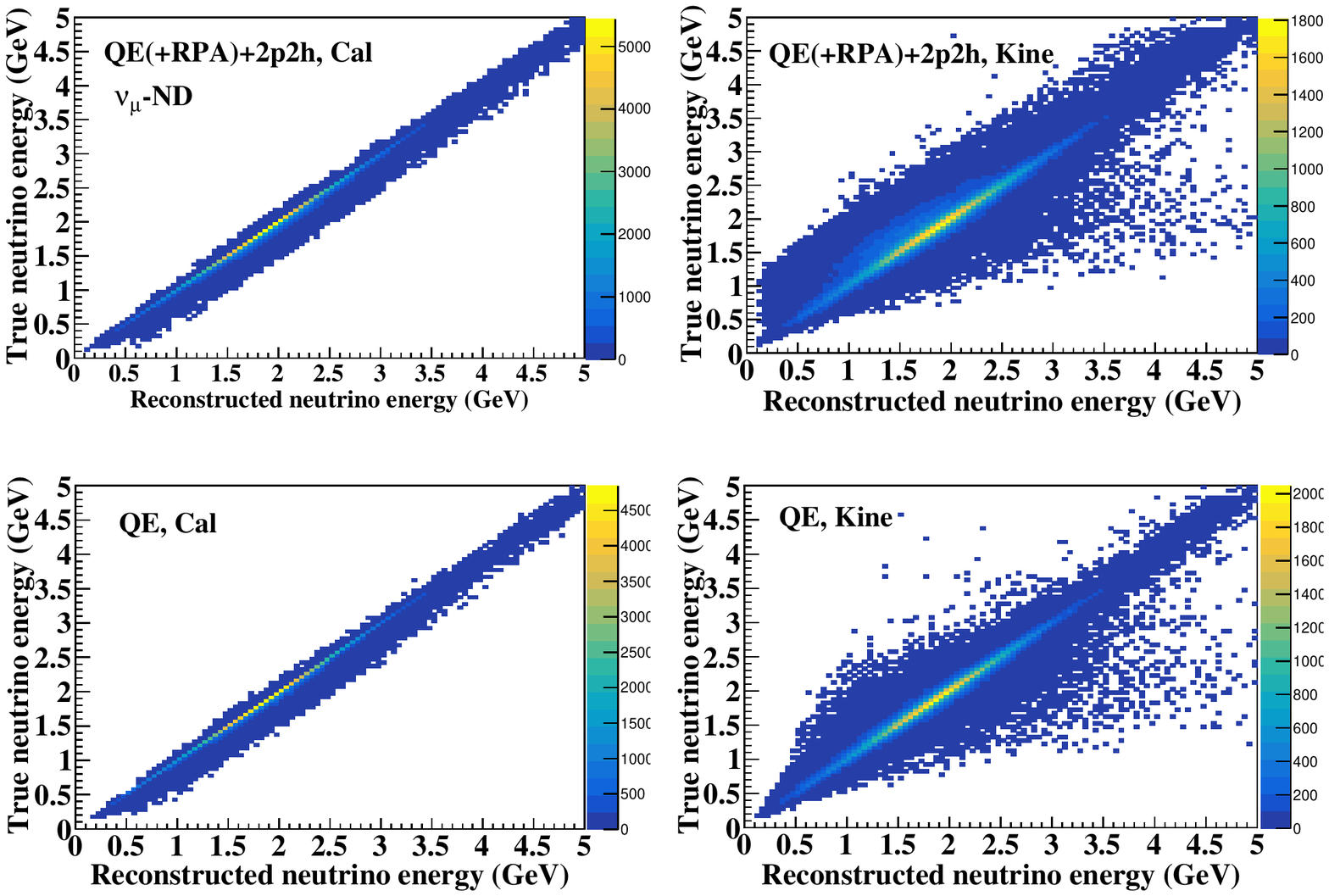}
\caption{Comparison of two-dimensional migration matrices using the calorimetric and kinematic reconstruction method for QE(+RPA)+2p2h and QE for GENIE using carbon as the target nucleus for ND.} 
\end{figure}

\begin{figure}[H] 
\centering\includegraphics[width=13.5cm, height=8cm]{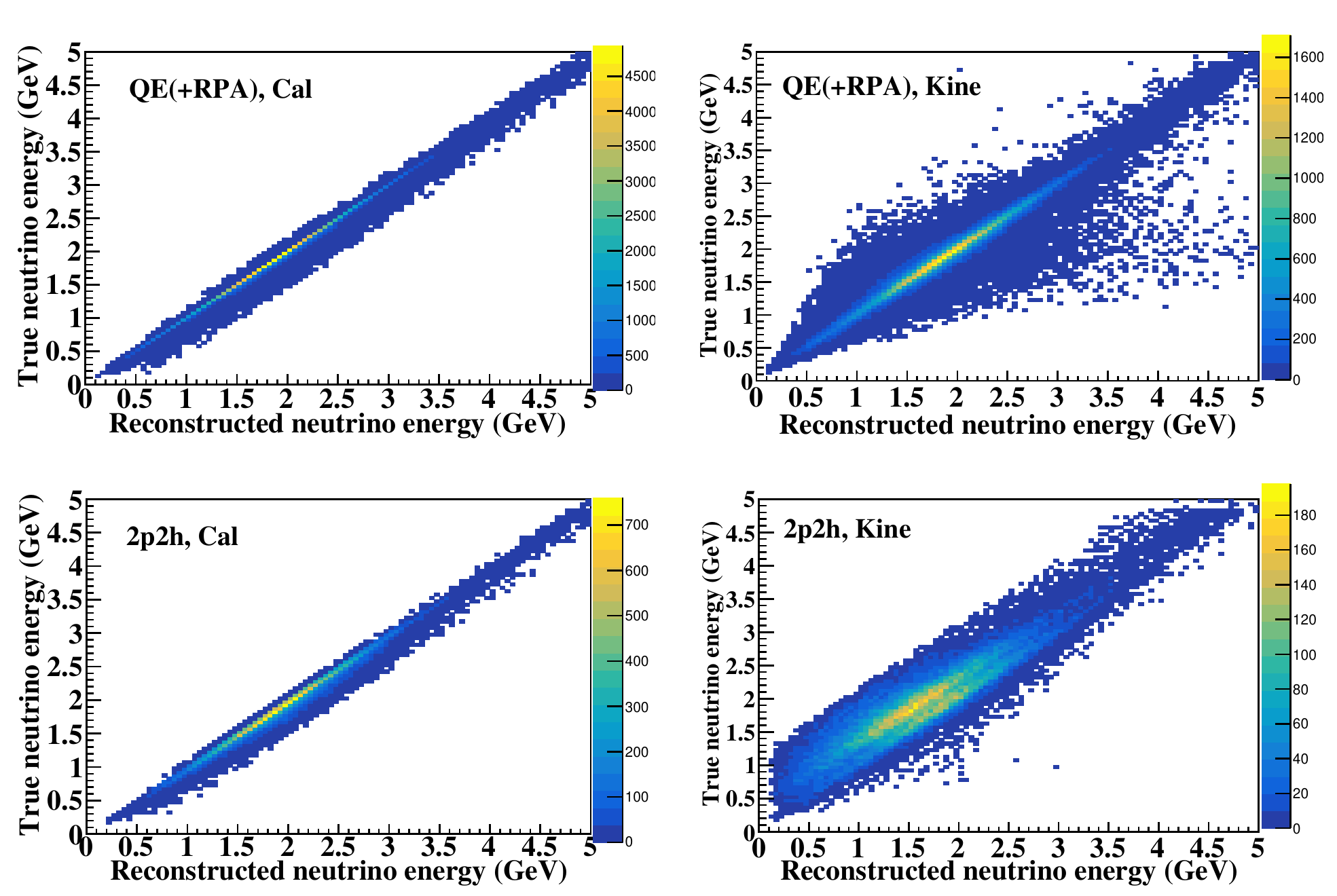}
\caption{Comparison of two-dimensional migration matrices using the calorimetric and kinematic reconstruction method for QE(+RPA) and 2p2h for GENIE using carbon as the target nucleus for ND.} 
\end{figure}

detectors are made from planes of extruded polyvinyl chloride (PVC) cells. The flavor composition of NuMI beam is 97.5\% $\nu_{\mu}$, 1.8\% $\bar\nu_{\mu}$ and 0.7\% $\nu_{e}+\bar\nu_{e}$. The primary physics goal of NO$\nu$A is to make precise measurements of $\theta_{23}$ mixing angle and its octant, neutrino mass hierarchy, and CP-violating phase $\delta_{CP}$ in the lepton sector. The $\nu_{\mu}(\bar\nu_{\mu})$ disappearance channel is used to measure $\Delta m^{2}_{32}$ and $\sin^{2}\theta_{23}$ while $\nu_{e}(\bar\nu_{e})$ appearance channel allows to measure mass hierarchy, $\theta_{13}$, $\theta_{23}$ and $\delta_{CP}$.

\begin{figure}[H] 
 \centering\includegraphics[width=13.5cm, height=8cm]{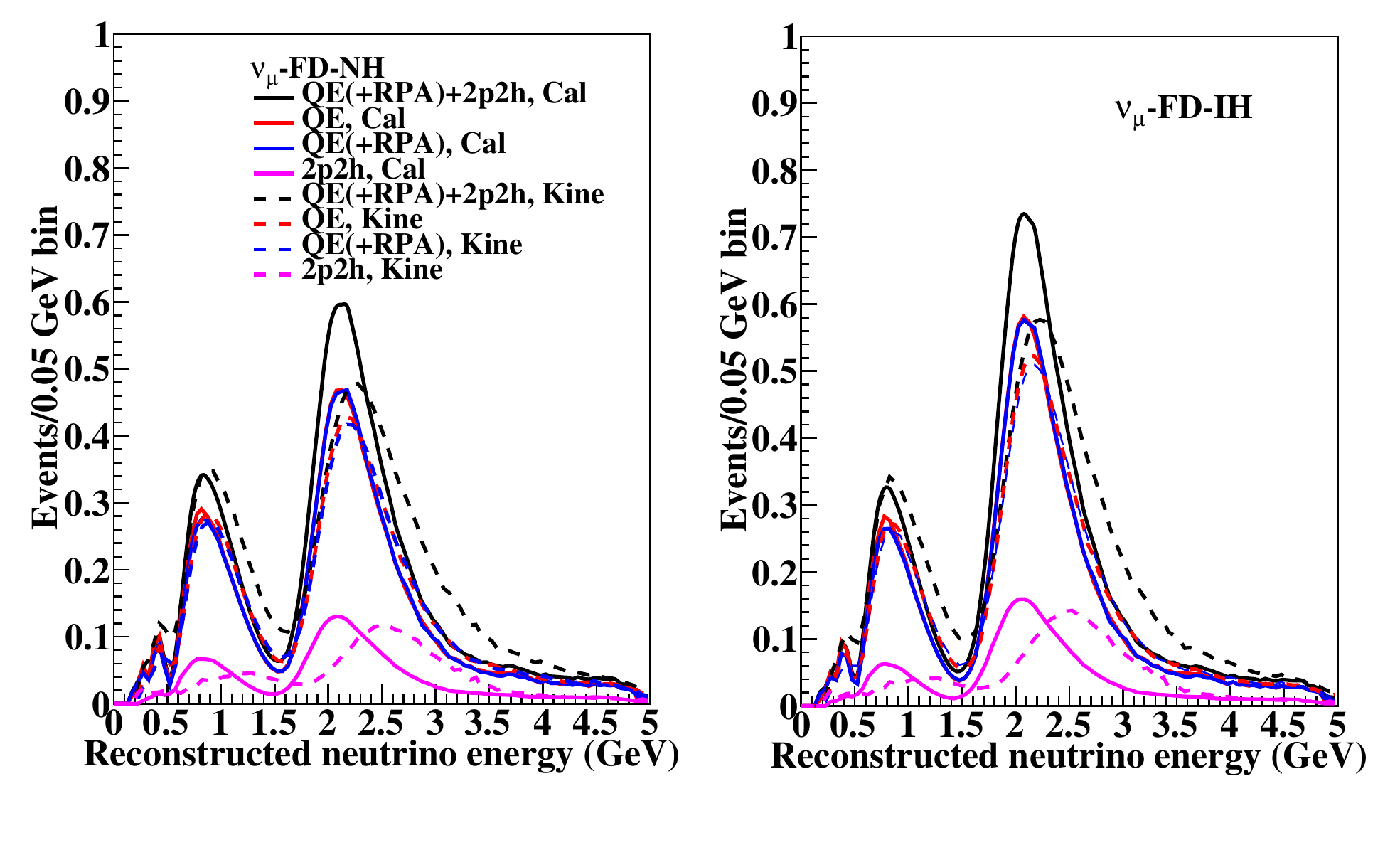}
\caption{Comparison of extrapolated FD events as a function of the calorimetric (solid line) and kinematic (dashed line) reconstruction energy for neutrino for NH (left panel) and IH (right panel) for different interactions: QE(+RPA)+2p2h (black line), QE (red line), QE(+RPA) (blue line) and 2p2h (magenta line).} 
\end{figure}

\begin{figure}[H] 
\centering\includegraphics[width=13.5cm, height=8cm]{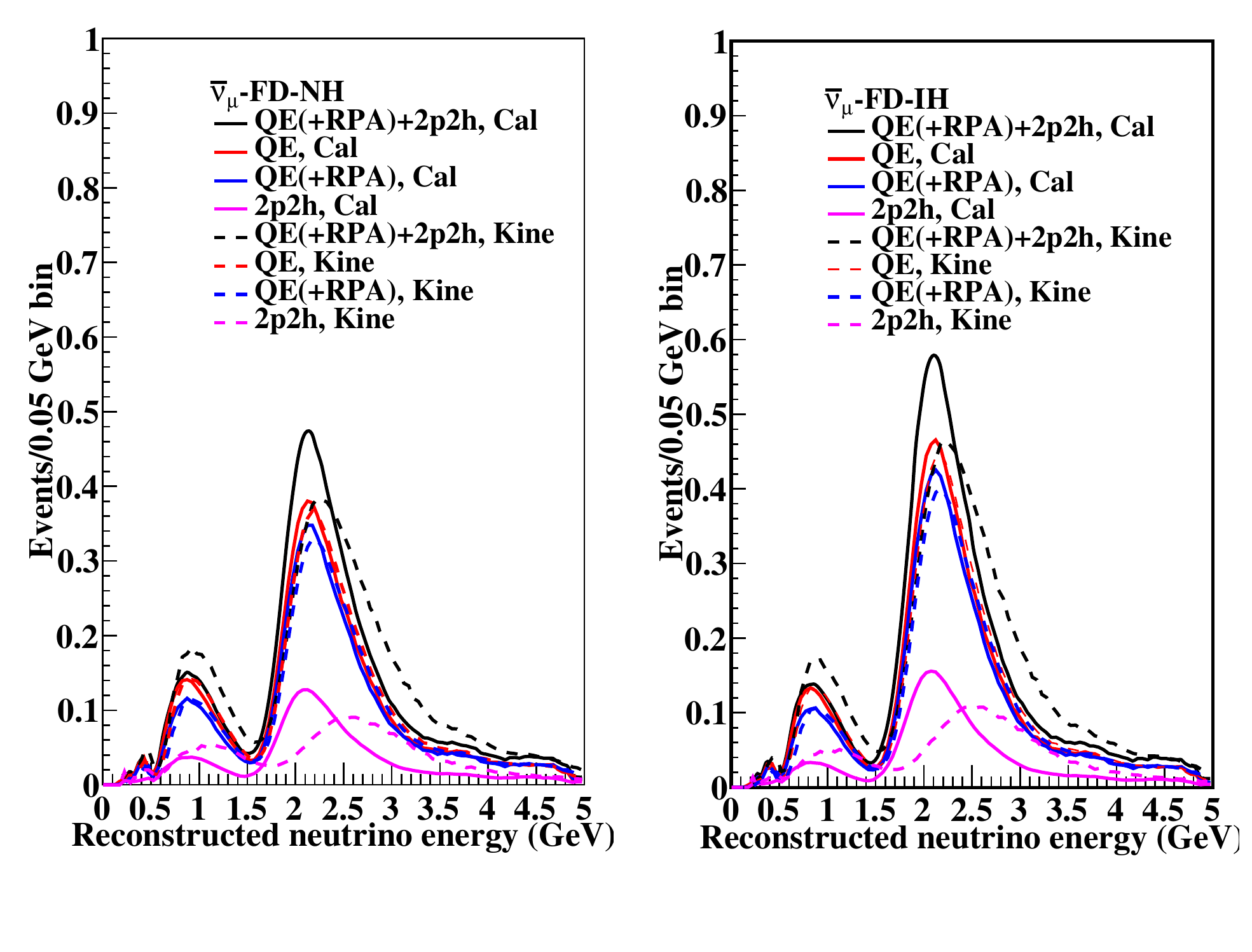}
\caption{Comparison of extrapolated FD events as a function of the calorimetric (solid line) and kinematic (dashed line) reconstruction energy for antineutrino for NH (left panel) and IH (right panel) for different interactions: QE(+RPA)+2p2h (black line), QE (red line), QE(+RPA) (blue line) and 2p2h (magenta line).} 
\end{figure}

\section{Energy Reconstruction Methods}
\label{sec:3}

This work aims to compare the two methods for reconstructing the neutrino energy-(i) the kinematic method, and (ii) the calorimetric method, with the inclusion of nuclear effects, and how they percolate into the sensitivity of neutrino oscillation parameters. The kinematic method is directly related to the kinematics (energy $E_{\mu}$ and angle $\theta_{\mu}$) of the outgoing lepton. It is based on the assumption of QE scattering when the nucleon is at rest, and constant binding energy is taken into account. The reconstructed neutrino energy with this approach is defined as \cite{Coloma:2013tba}, 

\begin{equation}
E^{kin}_{\nu}=\frac{ 2(M_{n}-E_{B})E_{\textit{l}}-(E^{2}_{B}-2M_{n}E_{B}+\Delta M^{2})}{2(M_{n}-E_{B}-E_{\textit{lep}}+\textit{p}_{\textit{lep}}\cos\theta_{\textit{lep}})}
\end{equation}\\

where $\Delta M^{2}=M^{2}_{n}-M^{2}_{p}+\textit{m}^{2}_{\textit{lep}}$ and $E_{B}$ is the average binding energy of the nucleon inside the nucleus.
$M_{n}$ represents the free neutron rest mass, $E_{\textit{lep}}$ and $\theta_{\textit{lep}}$ are the energy and angle of the outgoing lepton respectively. This equation is justified only for true QE interaction with the neutron at rest. Any mixture of other interactions gives incorrect reconstruction energy. For example, if a pion produced in the initial state is absorbed or undetected, the energy reconstruction for QE is lower than the true value by $\sim$ 300-350 MeV \cite{Ankowski:2015jya}. We use the outgoing lepton kinematics information also for non-QE events in the kinematic analysis.\\

Contrary to the kinematic method, in the calorimetric method, the incoming neutrino energy is reconstructed by adding the energies of all the observed particles in the final state, and no information is needed regarding the direction of the outgoing particles. It is computed as $E_{cal}$=$E_{\mu}+E_{had}$. Muon energy $E_{\mu}$ is reconstructed from the length of the muon track in the detector and the hadron energy $E_{had}$ is calculated from calorimetry by adding all the visible energy except the muonic energy. Thus, this method requires the information of all the detectable FSIs on an event-by-event basis and is therefore considered to be more precise than the kinematic method which is based on only the kinematics of the outgoing lepton. The calorimetric method is also applicable to all kinds of CC interactions. This method works effectively if it can reconstruct the hadrons precisely and there is no missing energy. Typically neutrons escape the detection and the undetected mesons lower the energy reconstruction by the value of the pion mass, at least $\sim$ 135 MeV. Thus, the calorimetric method mainly depends on the capability and design of the detector to accurately reconstruct the final state particles. And this justifies more impact of detector effects in the calorimetric method. Using the calorimetric approach, the energy of a CC neutrino, scattering off a nuclear target producing 'm' mesons and knocking out 'n' nucleons can be reconstructed as:

\begin{figure}[H] 
\centering\includegraphics[width=13.5cm, height=8cm]{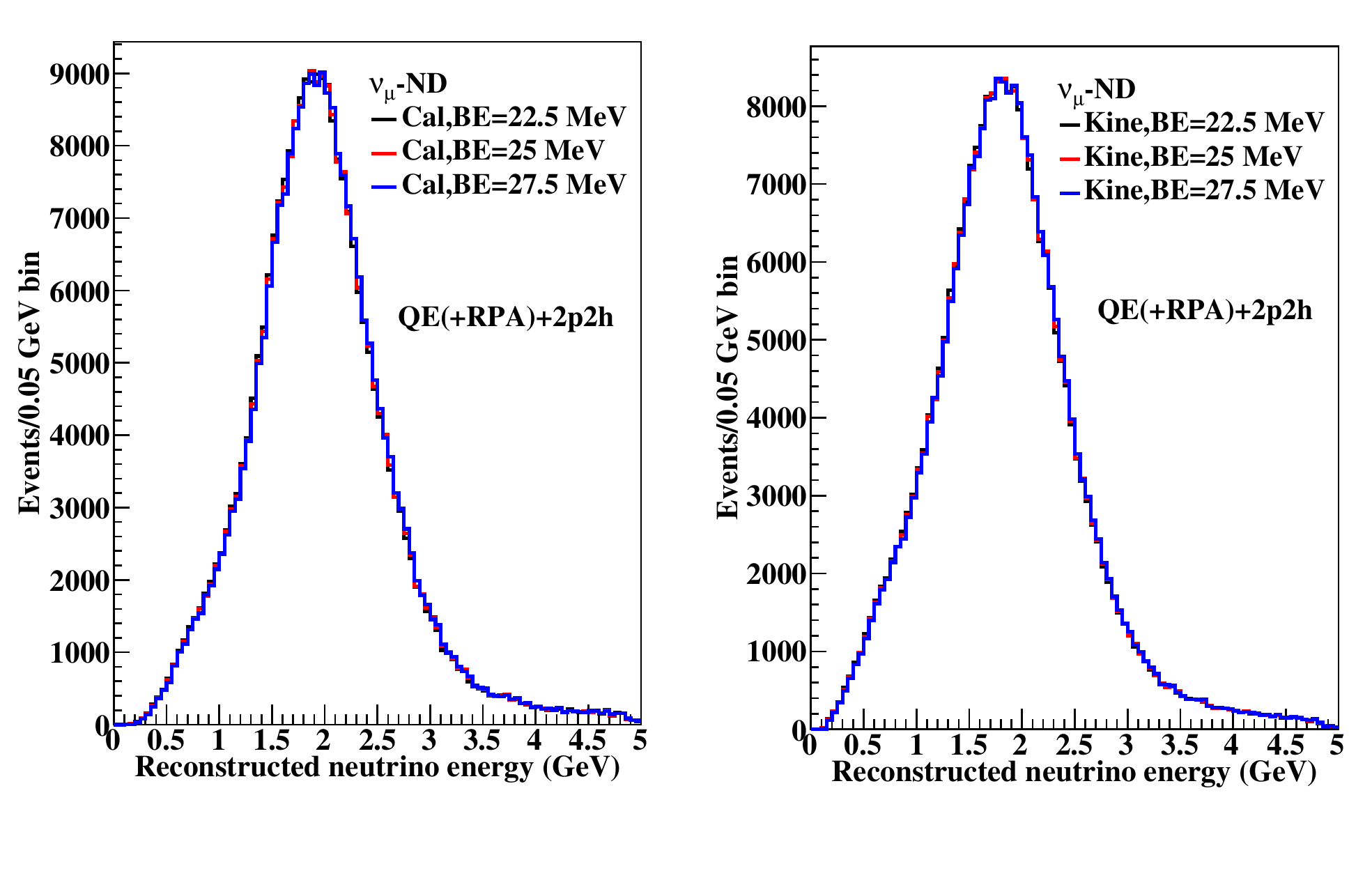}
\caption{Comparison of ND events for different binding energies as a function of the calorimetric (solid line) and kinematic (dashed line) reconstruction energy for neutrino for QE(+RPA)+2p2h process.} 
\end{figure}

\begin{equation}
E^{cal}_{\nu}=E_{lep}+\sum_{i}(E_{p}-M)+\epsilon_{n}+\sum_{m}E_{m}
\end{equation}

where $E_{lep}$ is the energy of the outgoing lepton, $E_{p}$ and $M$ denote the energy of the $i$-th knocked-out nucleon and mass of the nucleon (target nucleus). $\epsilon_{n}$ represents the single-nucleon separation energies of the outgoing nucleons and we keep it to a fixed value of 25 MeV for the carbon target for both neutrino and antineutrino for both the reconstruction methods. $E_{m}$ stands for the energy of the $m$-th produced meson. Keeping this in view, we studied the variation in the ND event distribution with binding energy, for its three randomly chosen values (taking an uncertainty of 10\%, i.e. 22.5 MeV, 25 MeV, and 27.5 MeV), for both the reconstruction methods.

\section {Mass hierarchy and CP sensitivity}
\label{sec:4} 

In LBL neutrino experiments, the appearance channel $\nu_{\mu}\rightarrow\nu_{e}$ is very sensitive in exploring the CP-violation effect which is the most challenging problem in neutrino physics today. The oscillation probability of $\nu_{\mu}\rightarrow\nu_{e}$ through matter in the standard three-flavor scenario and constant density approximation can be expressed as \cite{DUNE:2020jqi},\\

$P(\nu_{\mu}\rightarrow\nu_{e})=\sin^{2}\theta_{23}\sin^{2}2\theta_{13}\frac{\sin^{2}(\Delta_{31}-aL)}{(\Delta_{31}-aL)^{2}}\Delta^{2}_{31}+\sin2\theta_{23}\sin2\theta_{13}\sin2\theta_{12}\frac{\sin(\Delta_{31}-aL)}{(\Delta_{31}-aL)}\Delta_{31}$

\begin{equation}
\times \frac{\sin(aL)}{aL}\Delta_{21}\cos(\Delta_{31}+\delta_{CP})+\cos^{2}\theta_{23}\sin^{2}2\theta_{12}\frac{\sin^{2}(aL)}{(aL)^{2}}\Delta^{2}_{21}
\end{equation}

where $a=\frac{G_{F}N_{e}}{\sqrt{2}}\approx\pm\frac{1}{3500 km}\frac{\rho}{3.0 g/cm^{3}}$, $G_{F}$ is the Fermi constant, $N_{e}$ is the number density of electrons in Earth's crust, $ \Delta m^{2}_{ij}$ is the difference between mass squares of neutrinos of $i$-th and $j$-th family, $\Delta_{ij}=1.267 \Delta m^{2}_{ij} L/E_{\nu}$, L is the baseline in km (810 km for NO$\nu$A), and $E_{\nu}$ is the neutrino energy in GeV, $\rho$=2.848 $g cm^{-3}$. As the true value of $\delta_{CP}, $ is not known, the analysis is done by scanning all possible true values of $\delta_{CP}$ over the entire range $-\pi<\delta_{CP}<+\pi$ and comparing them with the CP conserving values i.e., 0 or $\pm\pi$. To detect the CP violation, the value of the CP phase must be different from 0 or $\pm\pi$. Therefore, the CP-violation sensitivity can be obtained at a given true value of $\delta_{CP}$, by minimizing $\chi^{2}$ at its fixed test values of 0 and $\pi$. To investigate CPV sensitivity, we compute and study two quantities,
 
\begin{equation}
\Delta \chi^{2}_{0}=\chi^{2}(\delta_{CP}=0)-\chi^{2}_{true}
\end{equation}
 
\begin{equation}
\Delta \chi^{2}_{\pi}=\chi^{2}(\delta_{CP}=\pi)-\chi^{2}_{true}
\end{equation}

and then we consider
 \begin{equation}
 \Delta \chi^{2}=min(\Delta \chi^{2}_{0}, \Delta \chi^{2}_{\pi})
 \end{equation}

The significance of CP-violation is obtained using $\sigma=\sqrt\Delta \chi^{2}$ and it is shown in Fig. 16 (left panel).\\

Another important topic of interest in the neutrino oscillation experiments is to understand the true nature of neutrino mass ordering, i.e., whether it is normal or inverted. The mass hierarchy sensitivity is calculated by assuming a normal (inverted) hierarchy to be a true hierarchy and comparing it to a test inverted (normal) hierarchy. Following formulae are used to determine the $\Delta \chi^{2}$ for mass hierarchy sensitivity:

\begin{equation}
\Delta \chi^{2}_{MH}=\chi^{2}_{IH}-\chi^{2}_{NH}
\end{equation}

\begin{equation}
\Delta \chi^{2}_{MH}=\chi^{2}_{NH}-\chi^{2}_{IH}
\end{equation}

The true values for the oscillation parameters are used from Table I and the test values of $\Delta m^{2}_{32}$, $\sin^{2}\theta_{23}$ and $\sin^{2}\theta_{13}$ are marginalized over their $3\sigma$ ranges in both the cases. The result is shown in Fig. 16 (right panel).\\

\begin{figure}[H] 
\centering\includegraphics[width=17cm, height=13cm]{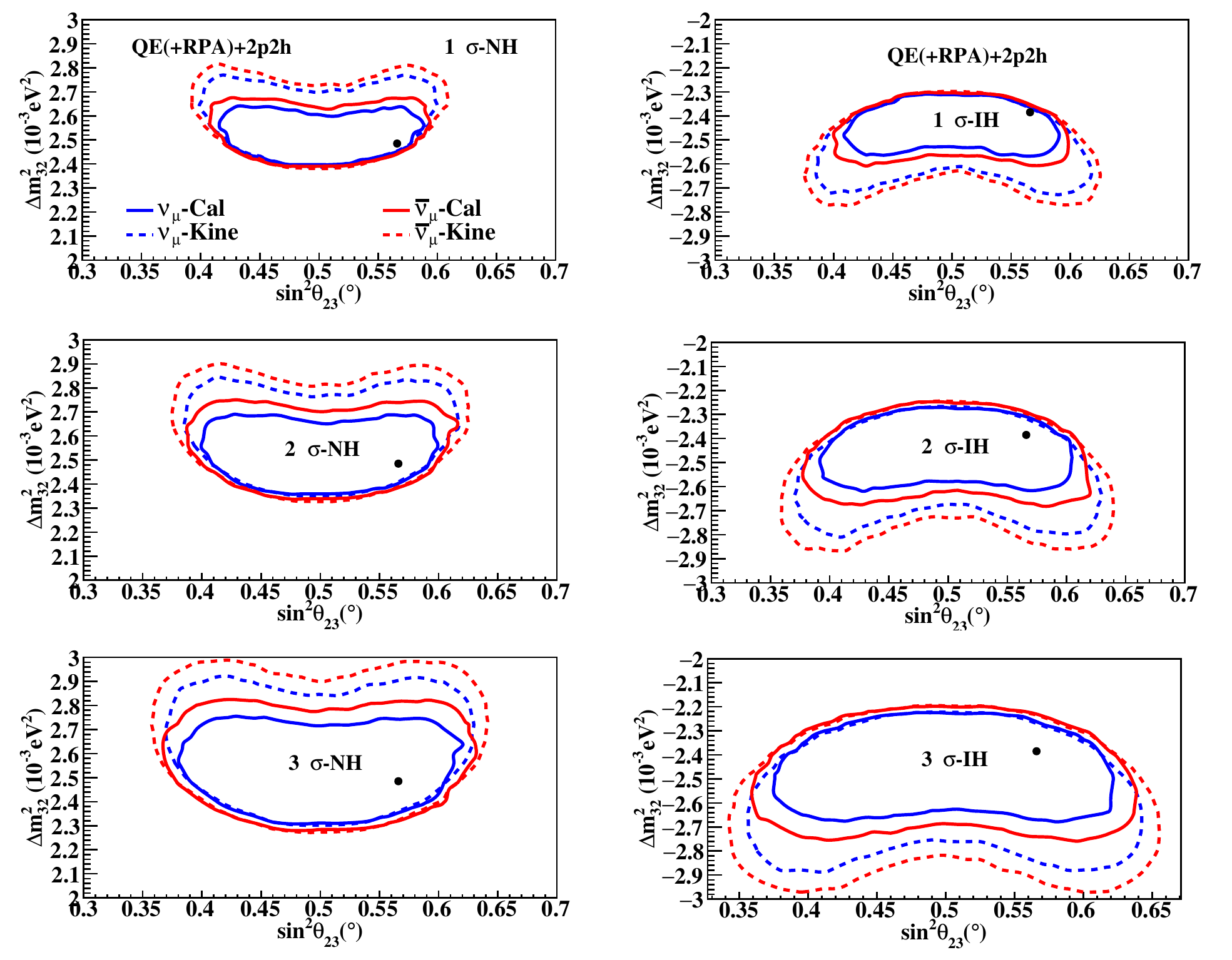}
\caption{Comparison of 1$\sigma$, 2$\sigma$ and 3$\sigma$ contours in $\Delta m^{2}_{32}$ vs $\sin^{2}\theta_{23}$ plane using the calorimetric (solid line) and kinematic methods (dashed line) of energy reconstruction for QE(+RPA)+2p2h for NH (left panel) and IH (right panel) for neutrino (blue line) and antineutrino (red line) without detector effect. The global best-fit point is shown by a black marker.} 
\end{figure}

\section{Physics and Simulation Details}
\label{sec:4}

In $\nu_{\mu}$ ($\bar\nu_{\mu})$ disappearance channel analysis, the $\nu_{\mu}$($\bar\nu_{\mu})$-CC interaction channel is used as the signal. In NO$\nu$A detectors, the $\nu_{\mu}$-CC interactions are identified from the long track of muons and any associated hadronic activity at the vertex. The un-oscillated spectra from the NuMI beam are first measured at the ND. Then these ND spectra are extrapolated to predict the FD spectra. The detailed extrapolation technique used in this work can be found in our previous work \cite{Deka:2021qnw}. For sensitivity analysis of neutrino oscillation parameters $\theta_{23}$ and $\Delta m^2_{32}$, we have used the Feldman-Cousins method \cite{Feldman:1997qc} to calculate the allowed confidence level in the parameter space \cite{Deka:2021qnw}. We have generated 1 million CC $\nu_{\mu}$ and $\bar\nu_{\mu}$ events sample for carbon target in the energy range 0-5 GeV using NO$\nu$A-ND neutrino and antineutrino flux. Simulation is done using neutrino event generator GENIE (Generates Events for Neutrino Interaction Experiments ) v3.0.6 \cite{Andreopoulos:2009rq}. Currently, GENIE is used by many neutrino baseline experiments, such as Miner$\nu$a \cite{MINERvA:2019ope}, MINOS \cite{Adamson:2007gu}, MicroBooNE \cite{Chen:2007ae}, T2K \cite{Abe:2018wpn} and NO$\nu$A experiments. 

We have considered four interaction processes - QE, the resonance from $\Delta$ resonant decay and contribution from higher resonances, nucleon-nucleon correlations 2p2h, and DIS (deep inelastic scattering) in both neutrino and antineutrino modes. We examined the simulated events with no pions in the final state. In GENIE, by default QE scattering is modeled using Llewellyn-Smith model \cite{LlewellynSmith:1971uhs} and the Local Fermi Gas (LFG) model \cite{Nieves:2004wx} is used as the nuclear model in our work. LFG model of nuclear ground state predicts a different initial nucleon momentum distribution than the RFG model \cite{NOvA:2020rbg}. When combined with the impact of Pauli suppression, this difference changes the available kinematic space in the QE interaction process. Resonance processes are predicted according to the formalism attributed to Berger and Sehgal model \cite{Berger:2007rq}. Further, to include the effect of long-range nuclear charge screening according to RPA correlations which also modify the kinematics of QE interaction, we simulate the QE interactions with the Nieves et. al. model \cite{Nieves:2004wx}. These effects suppress the QE interaction process significantly at low invariant four-momentum transferred to the nucleus ($Q^{2}$), and slightly enhance them at higher $Q^{2}$, relative to the RFG prediction. DIS interaction which is classified as a non-resonant process in GENIE is implemented following the method of Bodek and Yang \cite{Bodek:2002ps}. Using the Valencia model by Nieves et al. \cite{Nieves:2012yz}, two-nucleon knockout (2p2h) events are simulated. The relevant oscillation probability for $\nu_{\mu}\rightarrow\nu_{\mu}$ disappearance channel NO$\nu$A is calculated as shown in Ref. \cite{Deka:2021qnw}. Oscillation parameters best fit values are taken from \cite{deSalas:2020pgw} and shown in Table \ref{tab:data1}.\\

\begin{figure}[H] 
\centering\includegraphics[width=17cm, height=13cm]{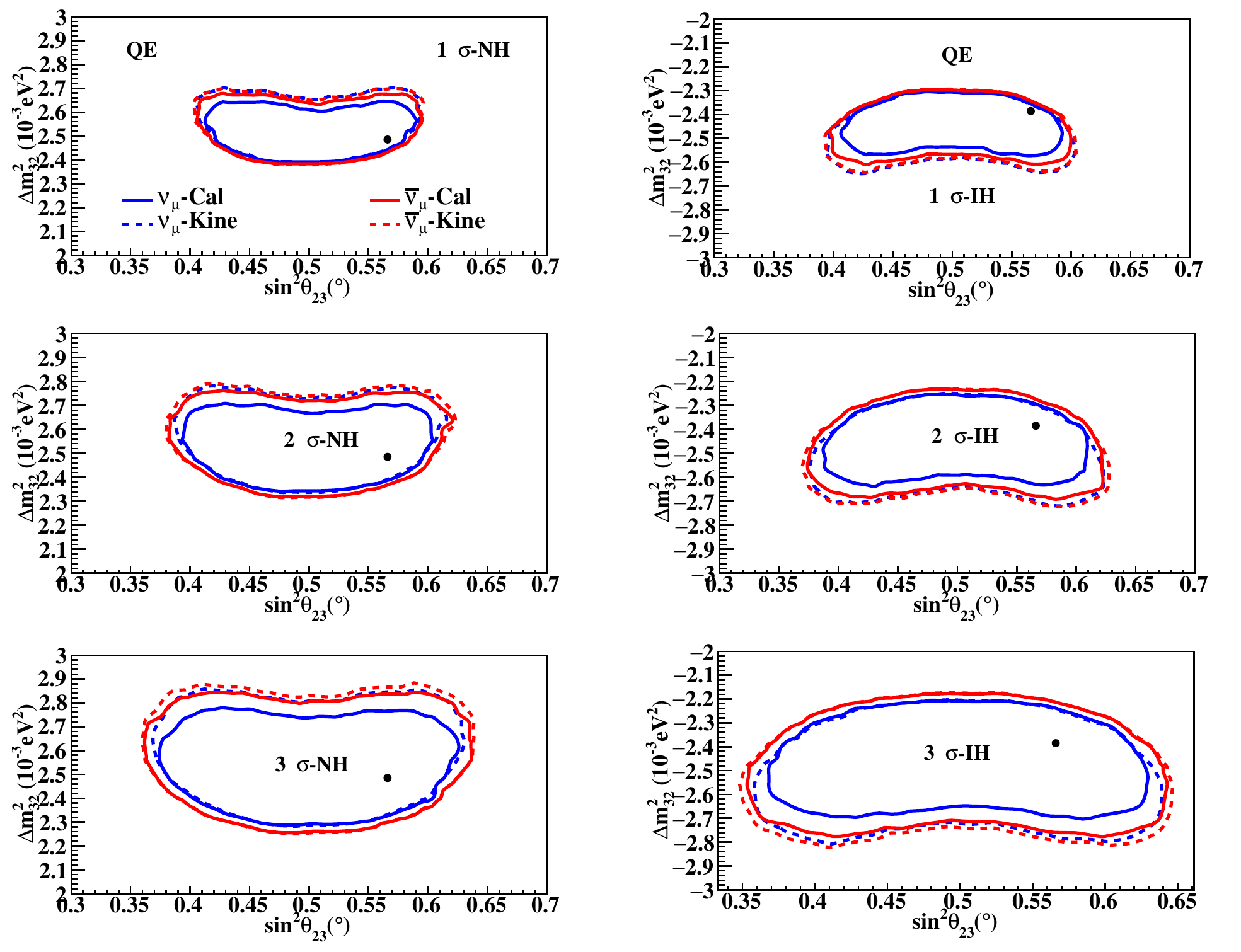}
\caption{Comparison of 1$\sigma$, 2$\sigma$ and 3$\sigma$ contours in $\Delta m^{2}_{32}$ vs $\sin^{2}\theta_{23}$ plane using the calorimetric (solid line) and kinematic methods (dashed line) of energy reconstruction for QE for NH (left panel) and IH (right panel) for neutrino (blue line) and antineutrino (red line) without detector effect.}
\end{figure}

Detector response and efficiency are also included in the analysis, where the measured energies are smeared with respect to the true values by finite detector resolution. Incorrect detector capabilities produce a non-vanishing probability such that an event with a true energy $E_{true}$ ends up being reconstructed with different energy $E_{rec}$. These non-vanishing probabilities are encoded in a set of migration matrices. We have considered realistic specification for NO$\nu$A - 31.2\% (33.9\%) efficiency of selection of $\nu_{\mu}$ ($\bar\nu_{\mu}$) events. For demonstration and comparison, we also analyzed for 100\% detector efficiency. 3.5\% muon energy resolution and 25\% hadron energy resolution are used respectively \cite{NOvA:2017ohq}, which gives an overall 7\% energy resolution for $\nu_{\mu}$-CC events for both detectors. We have simulated the true event rates with GENIE and global best fit values, and the fitted rates are obtained for both the reconstruction methods by varying the atmospheric oscillation parameters $\theta_{23}$ and $\Delta m^{2}_{32}$ in their 3$\sigma$ range. After determining the minimum $\chi^{2}$ value, $\chi^{2}_{best-fit}$, the confidence regions are found from the condition

\begin{equation}
\Delta \chi^{2}(\theta_{23}, \Delta m^{2}_{32}) \equiv \chi^{2}(\theta_{23}, \Delta m^{2}_{32})-\chi^{2}_{best-fit} < x
\end{equation}

where $x$= 2.30, 6.18, and 11.83 for the 1 $\sigma$, 2$\sigma$ and 3$\sigma$ confidence level, respectively.

\begin{table}[h]
\begin{center}
\begin{tabular}{|c|c|c|}
\hline
parameter & best fit & 3$\sigma$ range\\
\hline
$\Delta m_{21}^2[10^{-5} eV^{2}]$ & $7.50$ & $6.94-8.14$\\
$|\Delta m_{31}^2|[10^{-3} eV^{2}]$(NH) & $2.56$ & $2.46-2.65$ \\
$|\Delta m_{31}^2|[10^{-3} eV^{2}]$(IH) & $2.46$ & $2.37-2.55$ \\
$\sin^2\theta_{23}/10^{-1}$(NH) & $5.66$ & $4.46-6.09$ \\
$\sin^2\theta_{23}/10^{-1}$(IH) & $5.66$ & $4.41-6.09$ \\
$\sin^2\theta_{13}/10^{-2}$(NH) & $2.225$ & $2.015-2.417$ \\
$\sin^2\theta_{13}/10^{-2}$(IH) & $2.250$ & $2.039-2.441$ \\
$\sin^2\theta_{12}/10^{-1}$(IH) & $3.18$ & $2.71-3.70$ \\
\hline
\end{tabular}
\end{center}
\caption{3$\sigma$ values of neutrino oscillation parameters taken from \cite{deSalas:2020pgw}, which are used in this work.}
\label{tab:data1}
\end{table}

\begin{figure}[H] 
\centering\includegraphics[width=17cm, height=13cm]{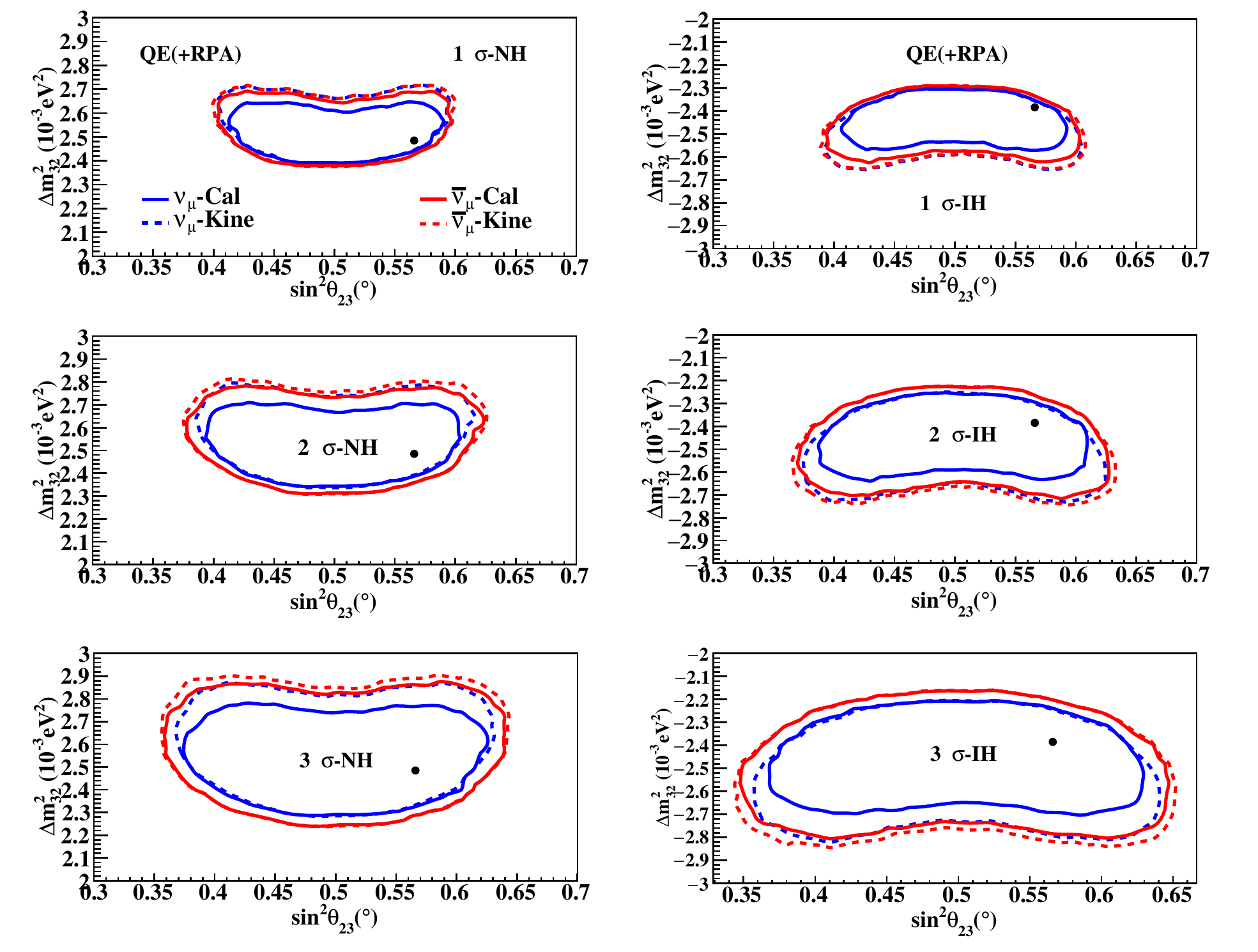}
\caption{Comparison of 1$\sigma$, 2$\sigma$ and 3$\sigma$ contours in $\Delta m^{2}_{32}$ vs $\sin^{2}\theta_{23}$ plane using the calorimetric (solid line) and kinematic methods (dashed line) of energy reconstruction for QE(+RPA) for NH (left panel) and IH (right panel) for neutrino (blue line) and antineutrino (red line) without detector effect}.
\end{figure}

\section{Results and Discussion}
\label{sec:5}

In Fig. 1, we have shown the total cross-sections per neutron on $^{12}C$ with no pions in the final state as a function of true neutrino energy for four different interaction processes- QE(+RPA)+2p2h (black line), QE only (red line), QE with RPA suppression QE(+RPA) (blue line), and 2p2h only (magenta line) for neutrino (solid line) and antineutrino (dashed line). As we are considering a complete model - QE with long-range RPA correlation and multi-nucleon enhancement 2p2h, therefore QE(+RPA)+2p2h interaction shows a higher contribution to the cross-section, which is also reflected in ND and FD event distribution. A suppression in cross-section due to RPA correction is observed in the energy range below 1 GeV for neutrino, after which both QE (without RPA) and QE with RPA curve tend to overlap with a slight suppression in cross-section for QE(+RPA). RPA correlations show maximum effect at forwarding angles where the contribution of squared four-momentum transfer $Q^{2}$ is small. They decrease with increasing angle and with increasing energy transfer. RPA correlations tend to lower the cross-section in the QE-peak region. Such effects have also been observed earlier too \cite{Nieves:2012yz, Singh:2019qac}.\\

 \begin{figure}[H] 
\centering\includegraphics[width=17cm, height=13cm]{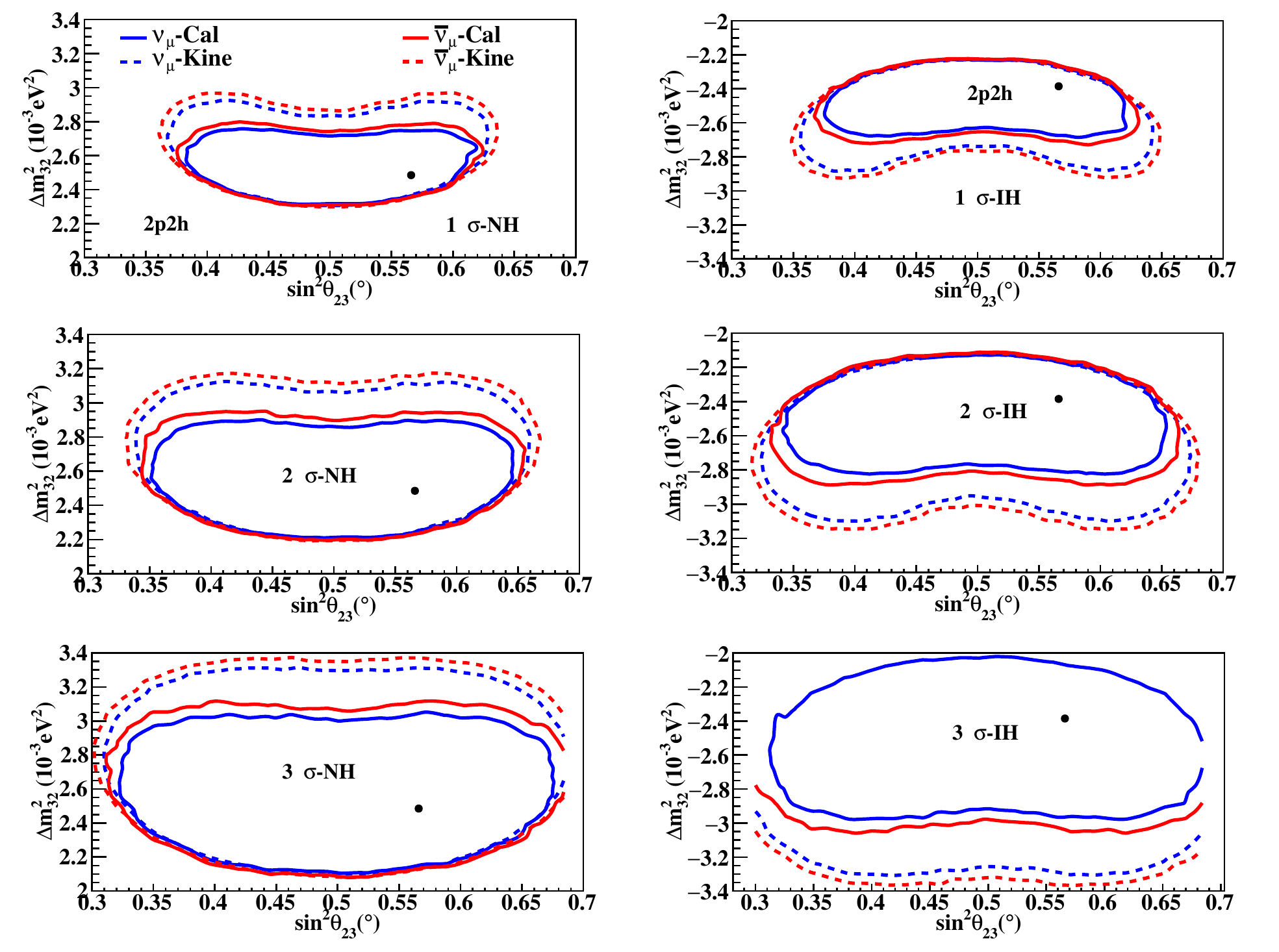}
\caption{Comparison of 1$\sigma$, 2$\sigma$ and 3$\sigma$ contours in $\Delta m^{2}_{32}$ vs $\sin^{2}\theta_{23}$ plane using the calorimetric (solid line) and kinematic methods (dashed line) of energy reconstruction for 2p2h only for NH (left panel) and IH (right panel) for neutrino (blue line) and antineutrino (red line) without detector effect.}
\end{figure}

In Fig. 2, the ND event distribution as a function of reconstructed energy, for two energy reconstruction methods are shown for different interactions - QE(+RPA)+2p2h (black line), QE only (red line), QE(+RPA) (blue line), and 2p2h only (magenta line). From the comparison of the two methods, we notice that event distribution with the kinematic method shifts towards the lower energy side compared to the calorimetric method. This shift is more for QE(+RPA)+2p2h and 2p2h while for QE and QE(+RPA), both the methods almost overlap in the case of neutrino. A similar observation is also made in the case of the antineutrino. The RPA suppression is less for neutrinos while it is distinctly visible in the antineutrino case. Also, the amplitude of the distribution is suppressed in the case of the calorimetric method (as compared to the kinematic method). These features will lead to shifts in sensitivity contours and can be attributed to different contributions in the energy reconstruction formulae in the two methods.\\

The importance of the differences between the migration matrices for calorimetric and kinematic reconstruction methods in the $\nu_{\mu}$ disappearance analysis for QE(+RPA)+2p2h, QE, QE(+RPA), and 2p2h interaction processes using carbon as the target nucleus for ND can be understood from Figs. 3 and 4. These matrices include FSI effects. Migration matrices are drawn between true and reconstructed neutrino energies where each element represents the probability that an event with specific true neutrino energy is reconstructed with different neutrino energy, which corresponds to an energy smearing as shown in Figs. 3 and 4. When the energy reconstruction method works well, the probability distribution is nearly symmetric about the line $E^{true}$=$E^{rec}$ and the probability of getting the energy reconstructed at a different value is relatively small which gives a diagonal line without any smearing. The broadening in the distribution comes from the Fermi motion of the bound nucleons \cite{Lalakulich:2012hs}. For all the events broadenings are observed to the line $E^{true}$=$E^{rec}$. From Figs. 3 and 4, we observe that the calorimetric method of energy reconstruction works rather well for all four processes in comparison with the kinematic method. The QE(+RPA)+2p2h, QE, QE(+RPA), and 2p2h events smear to a lesser extent when reconstructed with the calorimetric method as compared to the kinematic method. The kinematic method shows maximum smearing for QE(+RPA)+2p2h, QE, and QE(+RPA) which implies that the calorimetric method reconstructs incoming neutrino energy more precisely than the kinematic method. Smearing may be the effect of an incorrect description of nuclear effects. Hence, to reduce the systematic uncertainties in neutrino oscillation parameters, such errors in energy reconstruction in these interactions should be taken care of carefully.

\begin{figure}[H] 
\centering\includegraphics[width=17cm, height=12cm]{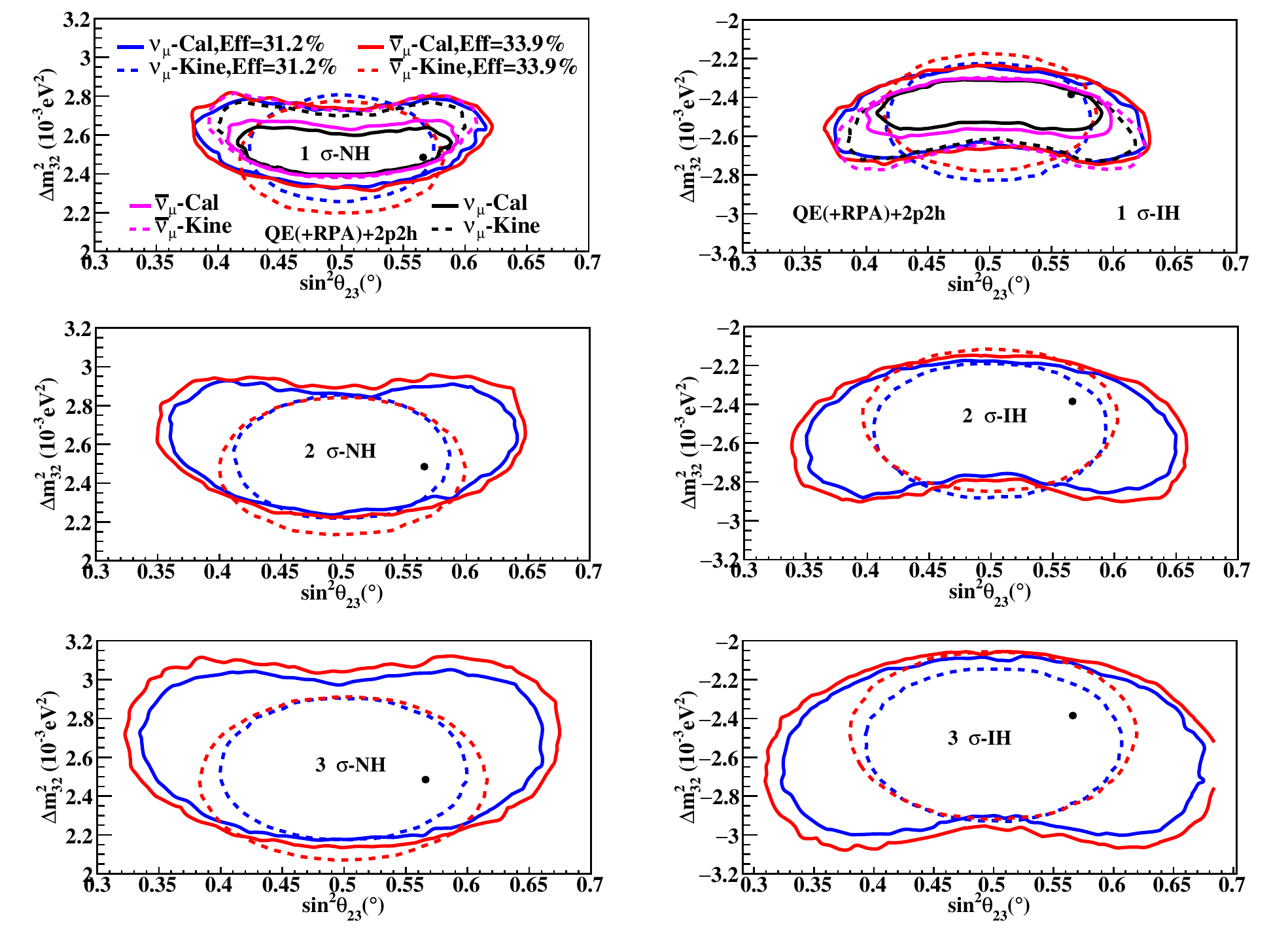}
\caption{Comparison of 1$\sigma$, 2$\sigma$ and 3$\sigma$ contours in $\Delta m^{2}_{32}$ vs $\sin^{2}\theta_{23}$ plane using the calorimetric (solid line) and kinematic methods (dashed line) of energy reconstruction for QE(+RPA)+2p2h for NH (left panel) and IH (right panel) with detector effect for detector efficiency 31.2\% ($\nu_{\mu}$) and 33.9\% ($\bar\nu_{\mu}$) for both neutrino (blue line) and antineutrino (red line). The black point indicates the global best fit value. The magenta and black curves are for no detector (ideal detector setup) case.} 
\end{figure}

In Figs. 5 and 6, we have presented the FD events generated by the extrapolation technique using migration matrices for both neutrino and antineutrino for both mass hierarchies. We observe that FD events distribution spectra broadens in the kinematic method while their amplitude increase in the calorimetric method. The broadening of spectra may be related to that of migration matrices in the kinematic reconstruction case (right panels in Figs. 3 and 4). Both the first and second oscillation maxima of the spectra with the kinematic method shift towards a little lower energy side than the calorimetric method. These spectra are the convolution of folding neutrino flux, cross-section (Fig. 1), ND event distribution (Fig. 2), migration matrices (Figs. 3 and 4), oscillation probability, nuclear effects, and detector response. Multi-nucleon effects and reconstruction methods change the amplitude as well as the position of the oscillation peaks in the FD events spectrum. For $\nu_{\mu}$ CC events, the position, and amplitude of the oscillation maximum in the FD energy spectra are strongly dependent on the neutrino oscillation parameters $\Delta m^{2}$ and $\theta_{23}$, therefore the differences in FD events distribution in the two construction methods (for all the interaction processes) are reflected in sensitivity analysis of these parameters.\\

How the choice of BE of target Carbon can affect event distribution in the two energy reconstruction methods is shown in Fig. 7, at ND of NO$\nu$A. However, almost no difference in a given reconstruction method is seen with a change in BE.\\

\begin{figure}[H] 
\centering\includegraphics[width=17cm, height=12cm]{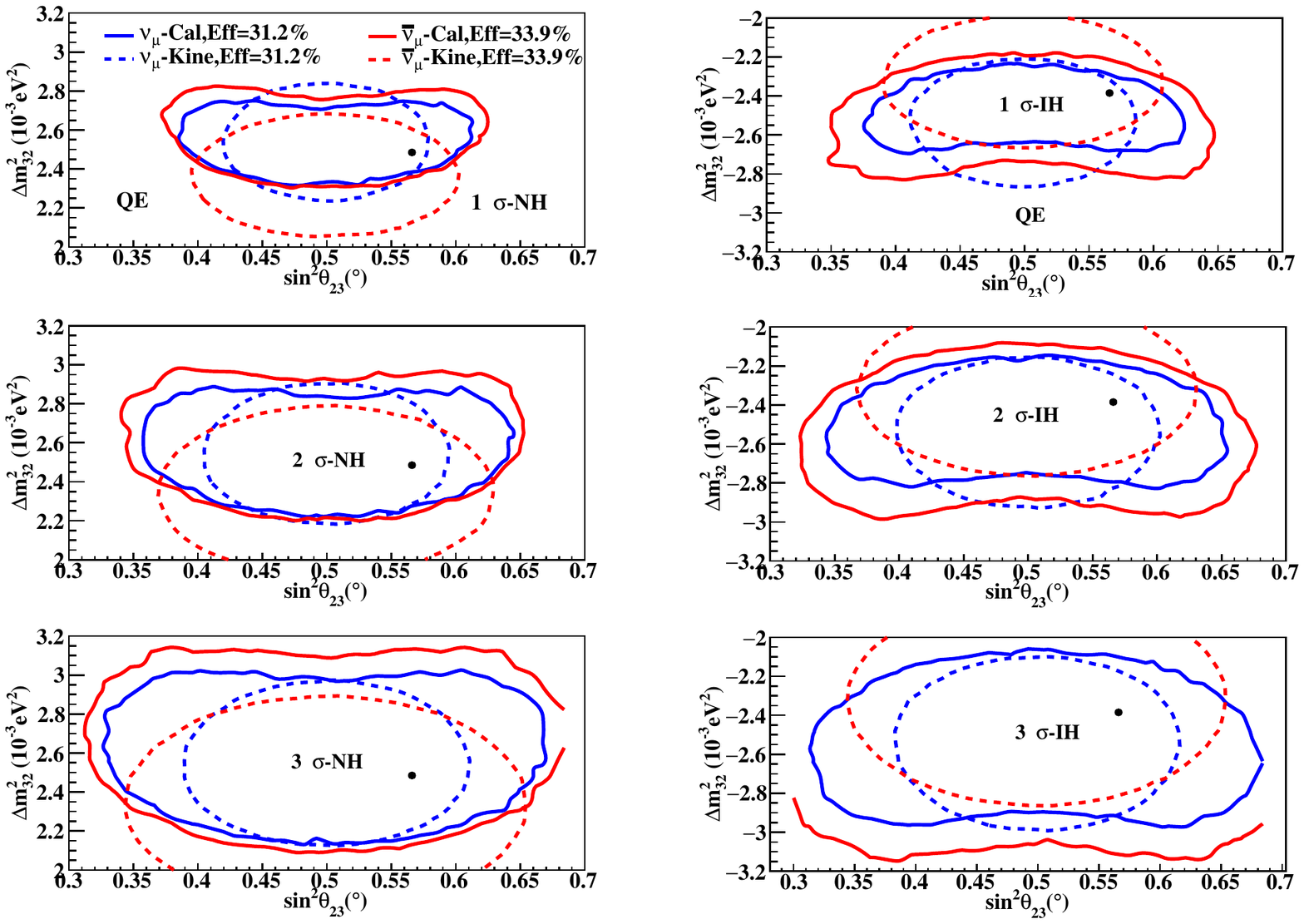}
\caption{Comparison of 1$\sigma$, 2$\sigma$ and 3$\sigma$ contours in $\Delta m^{2}_{32}$ vs $\sin^{2}\theta_{23}$ plane using the calorimetric (solid line) and kinematic methods (dashed line) of energy reconstruction for QE for NH (left panel) and IH (right panel) for neutrino (blue line) and antineutrino (red line) with detector effect for detector efficiency 31.2\% ($\nu_{\mu}$) and 33.9\% ($\bar\nu_{\mu}$). The black point indicates the global best fit value.} 
\end{figure}

In Figs. 8 and 9, we have shown the comparison of the calorimetric (solid line) and kinematic method (dashed line) of energy reconstruction in 1$\sigma$, 2$\sigma$ and 3$\sigma$ confidence regions in the ($\Delta m^{2}_{32}$, $\theta_{23}$) plane using the migration matrices for QE(+RPA)+2p2h (Fig. 8) and QE (Fig. 9) for NH (left panel) and IH (right panel) for both neutrino and antineutrino without detector effects. The difference between calorimetric and kinematic methods is distinctly visible for QE(+RPA)+2p2h interaction, for both the mass hierarchies and for both neutrino and antineutrino. However, in the case of QE interaction, the confidence regions for the kinematic method mostly coincide for neutrino and antineutrino in 1$\sigma$ for both NH and IH. The difference between the two reconstruction methods and the difference between neutrino and antineutrino is less for QE than QE(+RPA)+2p2h interaction. The contour area decreases significantly for the calorimetric method as compared to the kinematic method which would imply an improvement in the precision measurement of oscillation parameters using calorimetric reconstruction. Moreover, nuclear effects are not significant for the calorimetric method when detector effects are not included. \\

In Fig. 10, we have shown the comparison of the calorimetric (solid line) and kinematic (dashed line) method in 1$\sigma$, 2$\sigma$ and 3$\sigma$ contours in ($\Delta m^{2}_{32}, \sin^{2}\theta_{23}$) plane for QE(+RPA) for NH (left panel) and IH (right panel), for both neutrino and antineutrino (without detector effects). We observe that in QE interaction with RPA enhancement, the difference between calorimetric and kinematic methods is less for antineutrino than neutrino which is a reflection of the observations from Fig. 2 and 6. For the calorimetric method, the difference between neutrino and antineutrino confidence regions is seen to be significant, but for the kinematic method, they almost coincide with each other. A similar analysis for 2p2h is shown in Fig. 11, in which it is observed that the calorimetric method shows less uncertainty than the kinematic method.\\

\begin{figure}[H] 
\centering\includegraphics[width=17cm, height=12cm]{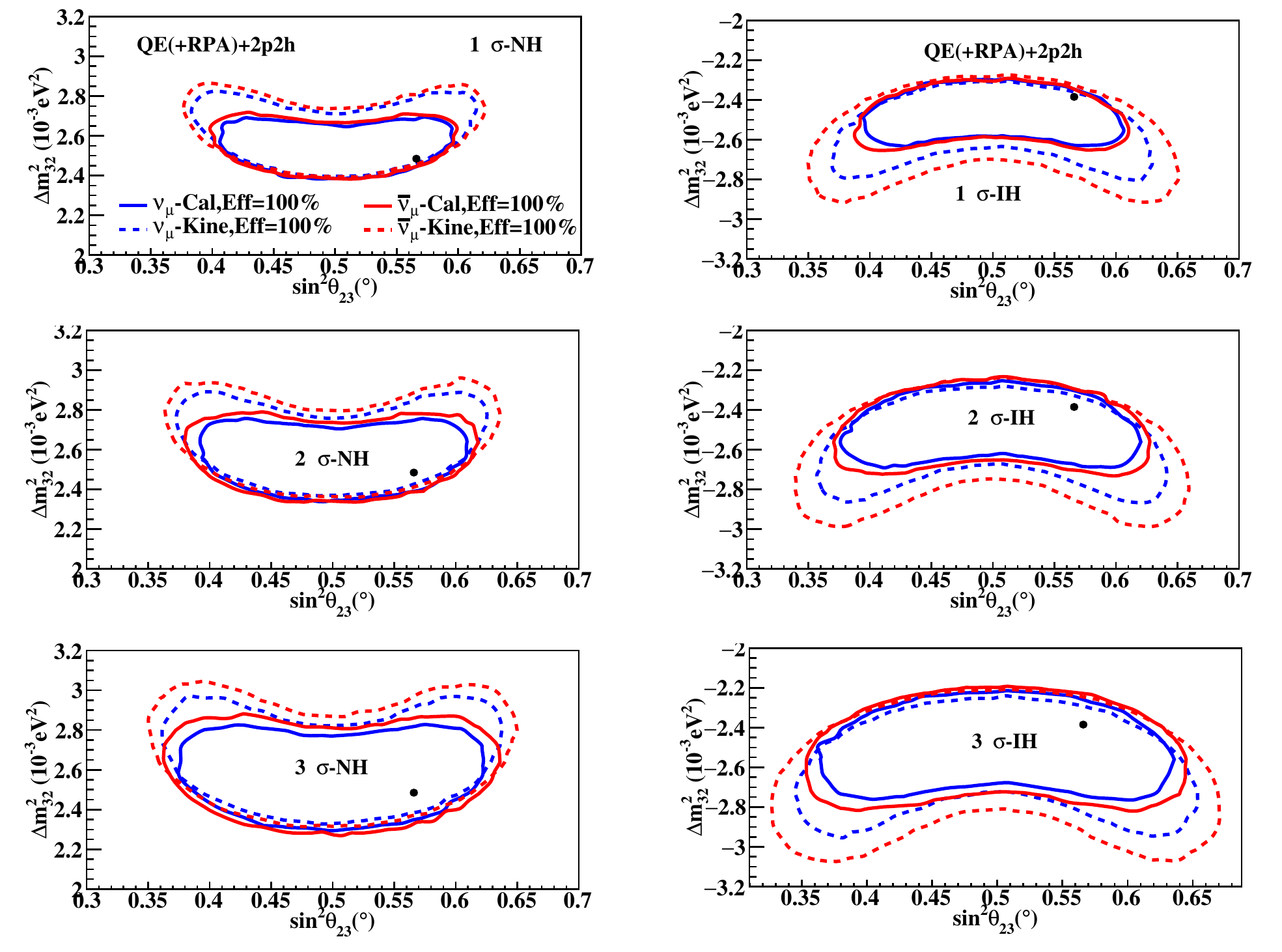}
\caption{Comparison of 1$\sigma$, 2$\sigma$ and 3$\sigma$ contours in $\Delta m^{2}_{32}$ vs $\sin^{2}\theta_{23}$ plane using the calorimetric (solid line) and kinematic methods (dashed line) of energy reconstruction for QE(+RPA)+2p2h for NH (left panel) and IH (right panel) for neutrino (blue line) and antineutrino (red line) with detector effect for detector efficiency 31.2\% ($\nu_{\mu}$) and 33.9\% ($\bar\nu_{\mu}$). The black point indicates the global best fit value.} 
\end{figure}

In Figs. 12 and 13, comparison of the calorimetric (solid line) and kinematic method (dashed line) of energy reconstruction in 1$\sigma$, 2$\sigma$ and 3$\sigma$ contours in $\Delta m^{2}_{32}$ vs $\sin^{2}\theta_{23}$ plane for QE(+RPA)+2p2h and QE for NH (left panel) and IH (right panel) for both neutrino and antineutrino with detector effects with efficiency 31.2\% and 33.9\% are shown. We have also shown the no detector effect in 1$\sigma$ contribution to estimate the deviation from no detector to detector effect. In Fig. 12, we observe that for both neutrino and antineutrino for efficiency 31.2\% and 33.9\%, the kinematic method shows better precision than the calorimetric method. A similar analysis for 100\% efficiency is shown in Fig. 14 and 15 respectively. These contours show how the confidence regions change when detector effects (along with nuclear effects) are included in the analysis. The quantum and nature of the above biases are almost the same in the two mass hierarchies for neutrino, and uncertainty is generally more in antineutrino. By comparison of respective figures, it is clear that in the no detector case (ideal detector) and the case with 100\% detector efficiency, areas inside the contour are less in the calorimetric method - and hence less uncertainty in parameter measurement with this method than with the kinematic method. On the other hand, in cases with detector efficiency 31.2\% and 33.9\%, the kinematic method shows less uncertainty.Thus, it is observed that the calorimetric method of energy reconstruction is more sensitive to detector effects when nuclear effects are included than the kinematic method. This can be justified as the calorimetric method depends more on the performance of the detector as energies of all particles in the final state need to be measured accurately. Detector effects reduce the uncertainty in the sensitivity analysis in both calorimetric and kinematic methods. This implies that improvement in detector efficiency is important for improving precision. We also observe that in the case of the kinematic method, the deviation from the no detector effect to the detector effect is less than in the case of the calorimetric method. This is because the kinematic method depends only on the kinematics of the outgoing lepton and in modern neutrino detectors muons are the most precisely reconstructed particles.\\

\begin{figure}[H] 
\centering\includegraphics[width=17cm, height=12cm]{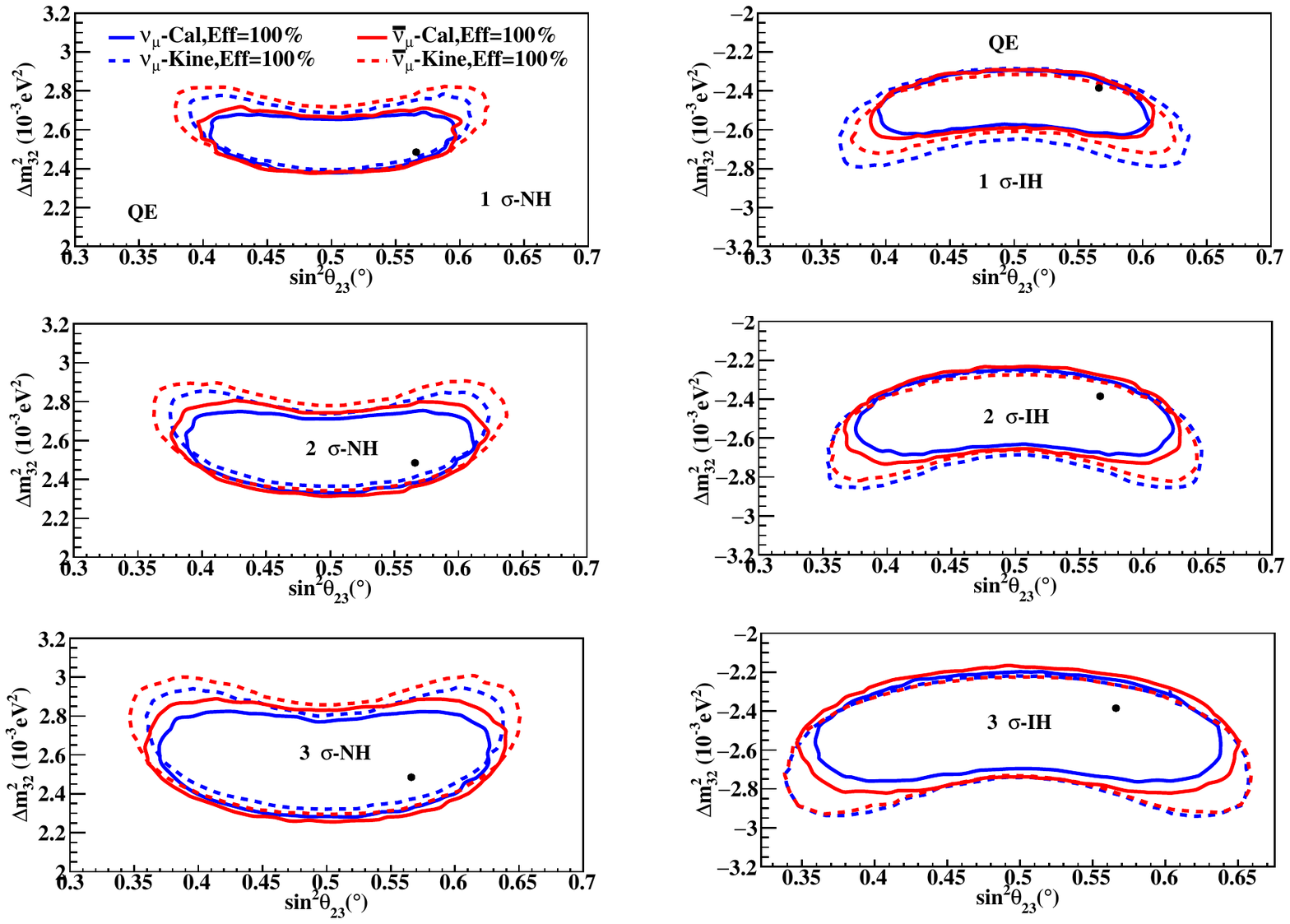}
\caption{Comparison of 1$\sigma$, 2$\sigma$ and 3$\sigma$ contours in $\Delta m^{2}_{32}$ vs $\sin^{2}\theta_{23}$ plane using the calorimetric (solid line) and kinematic methods (dashed line) of energy reconstruction for QE for NH (left panel) and IH (right panel) with detector effect for detector efficiency 100\% for both neutrino (blue line) and antineutrino (red line). The black point indicates the global best fit value.} 
\end{figure}

The dependence of sensitivity to the measurement of CPV phase $\delta_{CP}$, and of mass hierarchy significance, on energy reconstruction methods are shown in Fig. 16, for NH only for illustration, and here, detector effects have not been included. In the left panel, we show the CP sensitivity variation for the interaction processes QE(+RPA)+2p2h and QE, for neutrino+antineutrino mode. The 2$\sigma (1\sigma)$ line in the left panel corresponds to $\Delta \chi^{2}$= 4(1) for one degree of freedom (dof). This indicates a 95.45\%(68.3\%) probability of determining the CP-violation, and NH is considered as the true hierarchy. The variation of CP sensitivity with energy reconstruction methods is seen to be present at $\leqslant1\sigma$ level, and the difference between QE(+RPA)+2p2h and QE processes, for both the calorimetric and kinematic methods is seen to be of the order $\sim 0.2\sigma$ at $\delta_{CP}\sim -0.5\pi$ and $\delta_{CP}\sim 0.5\pi$. Though the sensitivity is more for the complete model, i.e. QE(+RPA)+2p2h than only the QE process, for a given interaction, however, the dependence on the energy reconstruction method is not very strong. In Fig 16 (right panel) we have shown the mass hierarchy significance plot with NH as the true hierarchy. The 2$\sigma$ and 1$\sigma$ lines in the right panel indicate approximately 95.45\% and 68.3\% probability of determining the MH correctly. The $\delta_{CP}$ values for which the curve is above 2$\sigma$ (3$\sigma$) are the values for which the hierarchy can be determined with 95.45\%(68.3\%) confidence level (CL). We observe the difference in CL in the two energy reconstruction methods is the maximum of the order $\sim 0.3\sigma$ at $\delta_{CP}\sim -0.5\pi$. It may be noted that qualitatively similar results on CP sensitivity MH significance have been reported earlier also \cite{Pershey:2018gtf, Singh:2019qac, Deepthi:2014iya}. Thus, energy reconstruction methods can play important role in MH measurements as well.\\

\begin{figure}[H] 
\centering\includegraphics[width=15cm, height=8cm]{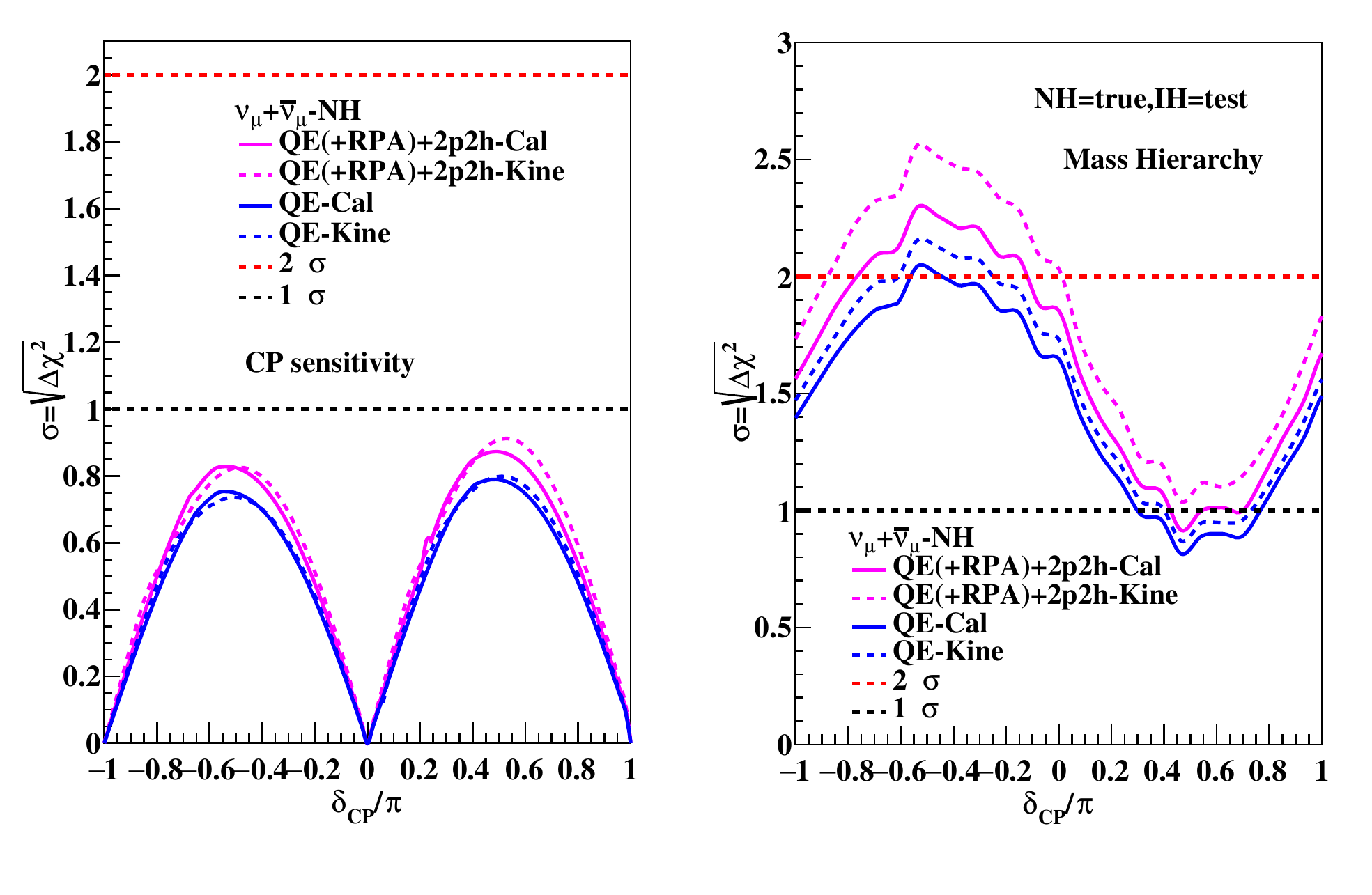}
\caption{Left panel: Plots for CP-violation sensitivity without detector effect for neutrino+antineutrino mode. Normal hierarchy (inverted hierarchy) is considered as a true hierarchy for QE(+RPA)+2p2h and QE interactions. Right panel: Mass hierarchy significance as a function of true $\delta_{CP}$ without detector effect for neutrino+antineutrino mode. Normal hierarchy (inverted hierarchy) is considered as true hierarchy and inverted (normal) is taken as test hierarchy for QE(+RPA)+2p2h and QE interactions.}
\end{figure}

\begin{table}[h]
\begin{center}
\begin{tabular}{|c|c|c|c|c|c|c|}
\hline
Details of the Process & $\sin^{2}\theta_{23,min}$ & $\Delta m^{2}_{32,min}[eV^{2}]$ & $\chi^{2}_{min}$ & $\sin^{2}\theta_{23},\Delta m^{2}_{32}$\scriptsize{(Global fit)} & Fig. no.\\
\hline
\scriptsize{$\nu_{\mu}$ QE(+RPA)+2p2h (Cal) (NH)} & \scriptsize{$0.524$} & \scriptsize{$2.48\times10^{-3}$} & \scriptsize{$2.57892$} & & \scriptsize{$(8)$}\\

\scriptsize{$\nu_{\mu}$ QE(+RPA)+2p2h (Kine) (NH)} & \scriptsize{$0.46$} & \scriptsize{$2.56\times10^{-3}$} & \scriptsize{$1.7667$} & & \scriptsize{$(8)$}\\

\scriptsize{$\nu_{\mu}$ QE (Cal) (NH)} & \scriptsize{$0.524$} & \scriptsize{$2.48\times10^{-3}$} & \scriptsize{$2.06898$} & & \scriptsize{$(9)$}\\

\scriptsize{$\nu_{\mu}$ QE (Kine) (NH)} & \scriptsize{$0.54$} & \scriptsize{$2.56\times10^{-3}$} & \scriptsize{$1.74289$} & & \scriptsize{$(9)$}\\

\scriptsize{$\nu_{\mu}$ QE(+RPA) (Cal) (NH)} & \scriptsize{$0.524$} & \scriptsize{$2.48\times10^{-3}$} & \scriptsize{$2.02843$} 	& \scriptsize{$0.566, 2.47\times10^{-3}$ (NH)} & \scriptsize{$(10)$}\\

\scriptsize{$\nu_{\mu}$ QE(+RPA) (Kine) (NH)} & \scriptsize{$0.54$} & \scriptsize{$2.56\times10^{-3}$} & \scriptsize{$1.62356$}	& & \scriptsize{$(10)$}\\

\scriptsize{$\nu_{\mu}$ 2p2h (Cal) (NH)} & \scriptsize{$0.524$} & \scriptsize{$2.48\times10^{-3}$} & \scriptsize{$0.55554$} 	& & \scriptsize{$(11)$}\\

\scriptsize{$\nu_{\mu}$ 2p2h (Kine) (NH)} & \scriptsize{$0.524$} & \scriptsize{$2.56\times10^{-3}$} & \scriptsize{$0.393395$} 	& & \scriptsize{$(11)$}\\

\hline
\scriptsize{$\nu_{\mu}$ QE(+RPA)+2p2h (Cal) (NH) (Eff=31.2\%)} & \scriptsize{$0.54$} & \scriptsize{$2.56\times10^{-3}$} & \scriptsize{$0.649116$} & & \scriptsize{$(12)$}\\

\scriptsize{$\nu_{\mu}$ QE(+RPA)+2p2h (Kine) (NH) (Eff=31.2\%)} & \scriptsize{$0.492$} & \scriptsize{$2.48\times10^{-3}$} & \scriptsize{$1.58913$} & & \scriptsize{$(12)$}\\

\scriptsize{$\nu_{\mu}$ QE (Cal) (NH) (Eff=31.2\%)} & \scriptsize{$0.476$} & \scriptsize{$2.48\times10^{-3}$} & \scriptsize{$0.580956$} & \scriptsize{$0.566, 2.47\times10^{-3}$ (NH)} & \scriptsize{$(13)$}\\

\scriptsize{$\nu_{\mu}$ QE (Kine) (NH) (Eff=31.2\%)} & \scriptsize{$0.492$} & \scriptsize{$2.56\times10^{-3}$} & \scriptsize{$1.47534$} & & \scriptsize{$(13)$}\\

\hline
\scriptsize{$\nu_{\mu}$ QE(+RPA)+2p2h (Cal) (NH) (Eff=100\%)} & \scriptsize{$0.54$} & \scriptsize{$2.56\times10^{-3}$} & \scriptsize{$2.23225$} & & \scriptsize{$(14)$}\\

\scriptsize{$\nu_{\mu}$ QE(+RPA)+2p2h (Kine) (NH) (Eff=100\%)} & \scriptsize{$0.508$} & \scriptsize{$2.56\times10^{-3}$} & \scriptsize{$1.48601$} & & \scriptsize{$(14)$}\\

\scriptsize{$\nu_{\mu}$ QE (Cal) (NH) (Eff=100\%)} & \scriptsize{$0.54$} & \scriptsize{$2.56\times10^{-3}$} & \scriptsize{$2.05989$} & \scriptsize{$0.566, 2.47\times10^{-3}$ (NH)} & \scriptsize{$(15)$}\\

\scriptsize{$\nu_{\mu}$ QE (Kine) (NH) (Eff=100\%)} & \scriptsize{$0.476$} & \scriptsize{$2.56\times10^{-3}$} & \scriptsize{$1.50631$} & & \scriptsize{$(15)$}\\

\hline
\hline

\scriptsize{$\nu_{\mu}$ QE(+RPA)+2p2h (Cal) (IH)} & \scriptsize{$0.476$} & \scriptsize{$2.4\times10^{-3}$} & \scriptsize{$2.68461$} & & \scriptsize{$(8)$}\\

\scriptsize{$\nu_{\mu}$ QE(+RPA)+2p2h (Kine) (IH)} & \scriptsize{$0.508$} & \scriptsize{$2.48\times10^{-3}$} & \scriptsize{$1.81933$} & & \scriptsize{$(8)$}\\

\scriptsize{$\nu_{\mu}$ QE (Cal) (IH)} & \scriptsize{$0.476$} & \scriptsize{$2.4\times10^{-3}$} & \scriptsize{$2.16304$} & & \scriptsize{$(9)$}\\

\scriptsize{$\nu_{\mu}$ QE (Kine) (IH)} & \scriptsize{$0.524$} & \scriptsize{$2.56\times10^{-3}$} & \scriptsize{$1.25265$} & & \scriptsize{$(9)$}\\

\scriptsize{$\nu_{\mu}$ QE(+RPA) (Cal) (IH)} & \scriptsize{$0.476$} & \scriptsize{$2.4\times10^{-3}$} & \scriptsize{$2.12139$} 	& & \scriptsize{$(10)$}\\

\scriptsize{$\nu_{\mu}$ QE(+RPA) (Kine) (IH)} & \scriptsize{$0.46$} & \scriptsize{$2.48\times10^{-3}$} & \scriptsize{$1.76127$} 	& \scriptsize{$0.566, 2.38\times10^{-3}$(IH)}& \scriptsize{$(10)$}\\

\scriptsize{$\nu_{\mu}$ 2p2h (Cal) (IH)} & \scriptsize{$0.524$} & \scriptsize{$2.4\times10^{-3}$} & \scriptsize{$	0.571004$} 	& & \scriptsize{$(11)$}\\
\scriptsize{$\nu_{\mu}$ 2p2h (Kine) (IH)} & \scriptsize{$0.54$} & \scriptsize{$2.48\times10^{-3}$} & \scriptsize{$0.435696$} 	& & \scriptsize{$(11)$}\\

\hline
\scriptsize{$\nu_{\mu}$ QE(+RPA)+2p2h (Cal) (IH) (Eff=31.2\%)} & \scriptsize{$0.492$} & \scriptsize{$2.4\times10^{-3}$} & \scriptsize{$0.695153$} & & \scriptsize{$(12)$}\\

\scriptsize{$\nu_{\mu}$ QE(+RPA)+2p2h (Kine) (IH) (Eff=31.2\%)} & \scriptsize{$0.492$} & \scriptsize{$2.48\times10^{-3}$} & \scriptsize{$1.4102$} & & \scriptsize{$(12)$}\\

\scriptsize{$\nu_{\mu}$ QE (Cal) (IH) (Eff=31.2\%)} & \scriptsize{$0.476$} & \scriptsize{$2.48\times10^{-3}$} & \scriptsize{$0.574578$} & \scriptsize{$0.566, 2.38\times10^{-3}$(IH)} & \scriptsize{$(13)$}\\

\scriptsize{$\nu_{\mu}$ QE (Kine) (IH) (Eff=31.2\%)} & \scriptsize{$0.508$} & \scriptsize{$2.48\times10^{-3}$} & \scriptsize{$1.05391$} & & \scriptsize{$(13)$}\\

\hline

\scriptsize{$\nu_{\mu}$ QE(+RPA)+2p2h (Cal) (IH) (Eff=100\%)} & \scriptsize{$0.508$} & \scriptsize{$2.4\times10^{-3}$} & \scriptsize{$2.32447$} & & \scriptsize{$(14)$}\\

\scriptsize{$\nu_{\mu}$ QE(+RPA)+2p2h (Kine) (IH) (Eff=100\%)} & \scriptsize{$0.46$} & \scriptsize{$2.48\times10^{-3}$} & \scriptsize{$1.48444$} & & \scriptsize{$(14)$}\\

\scriptsize{$\nu_{\mu}$ QE (Cal) (IH) (Eff=100\%)} & \scriptsize{$0.508$} & \scriptsize{$2.40\times10^{-3}$} & \scriptsize{$2.08782$} & \scriptsize{$0.566, 2.38\times10^{-3}$(IH)} & \scriptsize{$(15)$}\\

\scriptsize{$\nu_{\mu}$ QE (Kine) (IH) (Eff=100\%)} & \scriptsize{$0.476$} & \scriptsize{$2.48\times10^{-3}$} & \scriptsize{$1.55938$} & & \scriptsize{$(15)$}\\

\hline
\end{tabular}
\end{center}
\caption{Summary of best-fit values and corresponding $\chi^{2}$ values of the oscillation parameters along with global best-fit values for neutrino for different scenarios studied in this work.}
\label{tab:data2}
\end{table}

In Table \ref{tab:data2} and Table \ref{tab:data3}, we present the summary of best-fit values and corresponding $\chi^{2}$ values of the oscillation parameters, along with global best-fit values for neutrino and antineutrino for different scenarios studied in this work. We followed the Feldman-Cousins \cite{Feldman:1997qc} unified approach to determine confidence intervals within the range of likelihood ratios in pseudo experiments. From the results in these Tables, it is depicted how the best fit values change in different cases from no detector to detector effect, from NH to IH, and from calorimetric to kinematic methods for different interaction processes. The NO$\nu$A $\nu_{\mu}$ disappearance results constrains $\sin^{2}\theta_{23}$ around degenerate best fit points of 0.404 and 0.624. From NO$\nu$A result, it is observed that the best fit to the data gives $\Delta m^{2}_{32}=(+2.67\pm0.11)\times10^{-3}eV^{2}$ and $\sin^{2}\theta_{23}$ at the two statistically degenerate values $0.404^{+0.030}_{-0.022}$ and $0.624^{+0.022}_{-0.030}$ both at the 68\% C.L. in the NH. For IH, $\Delta m^{2}_{32}=(-2.72\pm0.11)\times10^{-3}eV^{2}$ and $\sin^{2}\theta_{23}=0.398^{+0.030}_{-0.022}$ or $0.618^{+0.022}_{-0.030}$ at 68\% C.L \cite{NOvA:2017ohq}. In NO$\nu$A, the statistical significance of these contours has been determined using the same Feldman-Cousins unified approach.

\section{Summary and Conclusion}
\label{sec:5}
Nuclear effects in neutrino interactions are one of the leading sources of systematic uncertainties in neutrino oscillation experiments. In recent T2K appearance results \cite{Abe:2019vii}, they are shown as among the largest contributors to the overall systematic errors. Also, neutrino energy reconstruction directly influences the precision of measurement of $\Delta m^{2}$ and $\theta_{23}$ parameters in LBL experiments. With this as motivation, in this work, we estimated the systematic uncertainties associated with the theoretical models of nuclear effects with two energy reconstruction methods - kinematic and calorimetric methods. We analyzed the impact of these two neutrino energy reconstruction methods on the extraction of neutrino oscillation parameters along with multi-nucleon enhancement in $\nu_{\mu}\rightarrow\nu_{\mu}$ disappearance channel, and CP sensitivity and mass hierarchy significance in $\nu_{\mu}\rightarrow\nu_{e}$ appearance channel of LBL NO$\nu$A experiment, in the energy range (0-5) GeV. We computed the contribution of RPA corrections and multi-nucleon effect 2p2h separately along with QE and the QE(+RPA)+2p2h interaction processes to neutrino-nucleus scattering cross-sections, ND event, and FD events, and migration matrices, using both the reconstruction methods. How detector efficiencies, resolutions, and nuclear model uncertainties affect the parameter oscillation analysis simulating the event distributions in FD using the realistic detector scenario for NO$\nu$A was also analyzed. The observed differences in the energy distributions of events both at ND and FD for the two energy reconstruction methods have important consequences for the oscillation analysis. The inaccuracies that appear in the description of nuclear effects are translated into a bias in the reconstruction of neutrino energy which further produces a bias in the extraction of neutrino oscillation parameters. The main results, i.e. the comparison of sensitivity analysis using the calorimetric and kinematic methods with the inclusion of nuclear effects in interactions and realistic detector effects of NO$\nu$A, were shown in Figs. 8-16.\\

\begin{table}[h]
\begin{center}
\begin{tabular}{|c|c|c|c|c|c|c|}
\hline

Details of the Process & $\sin^{2}\theta_{23,min}$ & $\Delta m^{2}_{32,min}[eV^{2}]$ & $\chi^{2}_{min}$ & $\sin^{2}\theta_{23},\Delta m^{2}_{32}$\scriptsize{(Global fit)} & Fig. no.\\
\hline

\scriptsize{$\bar\nu_{\mu}$ QE(+RPA)+2p2h (Cal) (NH)} & \scriptsize{$0.54$} & \scriptsize{$2.56\times10^{-3}$} & \scriptsize{$1.68726$} & & \scriptsize{$(8)$}\\

\scriptsize{$\bar\nu_{\mu}$ QE(+RPA)+2p2h (Kine) (NH)} & \scriptsize{$0.524$} & \scriptsize{$2.56\times10^{-3}$} & \scriptsize{$1.13777$} & & \scriptsize{$(8)$}\\

\scriptsize{$\bar\nu_{\mu}$ QE (Cal) (NH)} & \scriptsize{$0.54$} & \scriptsize{$2.56\times10^{-3}$} & \scriptsize{$1.45461$} & & \scriptsize{$(9)$}\\

\scriptsize{$\bar\nu_{\mu}$ QE (Kine) (NH)} & \scriptsize{$0.54$} & \scriptsize{$2.56\times10^{-3}$} & \scriptsize{$1.29659$} & & \scriptsize{$(9)$}\\

\scriptsize{$\bar\nu_{\mu}$ QE(+RPA) (Cal) (NH)} & \scriptsize{$0.54$} & \scriptsize{$2.56\times10^{-3}$} & \scriptsize{$1.25655$} 	& \scriptsize{$0.566, 2.47\times10^{-3}$ (NH)} & \scriptsize{$(10)$}\\

\scriptsize{$\bar\nu_{\mu}$ QE(+RPA) (Kine) (NH)} & \scriptsize{$0.54$} & \scriptsize{$2.56\times10^{-3}$} & \scriptsize{$1.11804$} 	& & \scriptsize{$(10)$}\\

\scriptsize{$\bar\nu_{\mu}$ 2p2h (Cal) (NH)} & \scriptsize{$0.54$} & \scriptsize{$2.56\times10^{-3}$} & \scriptsize{$0.439069$} 	& & \scriptsize{$(11)$}\\

\scriptsize{$\bar\nu_{\mu}$ 2p2h (Kine) (NH)} & \scriptsize{$0.524$} & \scriptsize{$2.56\times10^{-3}$} & \scriptsize{$0.287889$} 	& & \scriptsize{$(11)$}\\

\hline

\scriptsize{$\bar\nu_{\mu}$ QE(+RPA)+2p2h (Cal) (NH) (Eff=33.9\%)} & \scriptsize{$0.492$} & \scriptsize{$2.56\times10^{-3}$} & \scriptsize{$0.564969$} & & \scriptsize{$(12)$}\\

\scriptsize{$\bar\nu_{\mu}$ QE(+RPA)+2p2h (Kine) (NH) (Eff=33.9\%)} & \scriptsize{$0.492$} & \scriptsize{$2.48\times10^{-3}$} & \scriptsize{$0.940515$} & & \scriptsize{$(12)$}\\

\scriptsize{$\bar\nu_{\mu}$ QE (Cal) (NH) (Eff=33.9\%)} & \scriptsize{$0.524$} & \scriptsize{$2.56\times10^{-3}$} & \scriptsize{$0.433007$} & \scriptsize{$0.566, 2.47\times10^{-3}$ (NH)} & \scriptsize{$(13)$}\\

\scriptsize{$\bar\nu_{\mu}$ QE (Kine) (NH) (Eff=33.9\%)} & \scriptsize{$0.492$} & \scriptsize{$2.48\times10^{-3}$} & \scriptsize{$1.24153$} & & \scriptsize{$(13)$}\\

\hline

\scriptsize{$\bar\nu_{\mu}$ QE(+RPA)+2p2h (Cal) (NH) (Eff=100\%)} & \scriptsize{$0.54$} & \scriptsize{$2.56\times10^{-3}$} & \scriptsize{$1.41253$} & & \scriptsize{$(14)$}\\

\scriptsize{$\bar\nu_{\mu}$ QE(+RPA)+2p2h (Kine) (NH) (Eff=100\%)} & \scriptsize{$0.476$} & \scriptsize{$2.56\times10^{-3}$} & \scriptsize{$0.907663$} & & \scriptsize{$(14)$}\\

\scriptsize{$\bar\nu_{\mu}$ QE (Cal) (NH) (Eff=100\%)} & \scriptsize{$0.46$} & \scriptsize{$2.56\times10^{-3}$} & \scriptsize{$1.3784$} & \scriptsize{$0.566, 2.47\times10^{-3}$ (NH)} & \scriptsize{$(15)$}\\

\scriptsize{$\bar\nu_{\mu}$ QE (Kine) (NH) (Eff=100\%)} & \scriptsize{$0.476$} & \scriptsize{$2.56\times10^{-3}$} & \scriptsize{$1.00896$} & & \scriptsize{$(15)$}\\

\hline
\hline

\scriptsize{$\bar\nu_{\mu}$ QE(+RPA)+2p2h (Cal) (IH)} & \scriptsize{$0.46$} & \scriptsize{$2.48\times10^{-3}$} & \scriptsize{$1.88138$} & & \scriptsize{$(8)$}\\

\scriptsize{$\bar\nu_{\mu}$ QE(+RPA)+2p2h (Kine) (IH)} & \scriptsize{$0.524$} & \scriptsize{$2.56\times10^{-3}$} & \scriptsize{$1.16509$} & & \scriptsize{$(8)$}\\

\scriptsize{$\bar\nu_{\mu}$ QE (Cal) (IH)} & \scriptsize{$0.524$} & \scriptsize{$2.4\times10^{-3}$} & \scriptsize{$1.56191$} & & \scriptsize{$(9)$}\\

\scriptsize{$\bar\nu_{\mu}$ QE (Kine) (IH)} & \scriptsize{$0.476$} & \scriptsize{$2.4\times10^{-3}$} & \scriptsize{$1.47151$} & & \scriptsize{$(9)$}\\

\scriptsize{$\bar\nu_{\mu}$ QE(+RPA) (Cal) (IH)} & \scriptsize{$0.46$} & \scriptsize{$2.48\times10^{-3}$} & \scriptsize{$1.40621$} 	& \scriptsize{$0.566, 2.38\times10^{-3}$(IH)} & \scriptsize{$(10)$}\\

\scriptsize{$\bar\nu_{\mu}$ QE(+RPA) (Kine) (IH)} & \scriptsize{$0.54$} & \scriptsize{$2.48\times10^{-3}$} & \scriptsize{$1.31729$} 	& & \scriptsize{$(10)$}\\

\scriptsize{$\bar\nu_{\mu}$ 2p2h (Cal) (IH)} & \scriptsize{$0.46$} & \scriptsize{$2.48\times10^{-3}$} & \scriptsize{$0.477686$} 	& & \scriptsize{$(11)$}\\

\scriptsize{$\bar\nu_{\mu}$ 2p2h (Kine) (IH)} & \scriptsize{$0.46$} & \scriptsize{$2.48\times10^{-3}$} & \scriptsize{$0.322048$} 	& & \scriptsize{$(11)$}\\

\hline
\scriptsize{$\bar\nu_{\mu}$ QE(+RPA)+2p2h (Cal) (IH) (Eff=33.9\%)} & \scriptsize{$0.46$} & \scriptsize{$2.48\times10^{-3}$} & \scriptsize{$0.500182$} & & \scriptsize{$(12)$}\\

\scriptsize{$\bar\nu_{\mu}$ QE(+RPA)+2p2h (Kine) (IH) (Eff=33.9\%)} & \scriptsize{$0.492$} & \scriptsize{$2.48\times10^{-3}$} & \scriptsize{$0.977105$} & \scriptsize{$0.566, 2.38\times10^{-3}$(IH)} & \scriptsize{$(12)$}\\

\scriptsize{$\bar\nu_{\mu}$ QE (Cal) (IH) (Eff=33.9\%)} & \scriptsize{$0.476$} & \scriptsize{$2.56\times10^{-3}$} & \scriptsize{$0.481189$} & & \scriptsize{$(13)$}\\

\scriptsize{$\bar\nu_{\mu}$ QE (Kine) (IH) (Eff=33.9\%)} & \scriptsize{$0.492$} & \scriptsize{$2.32\times10^{-3}$} & \scriptsize{$1.15407$} & & \scriptsize{$(13)$}\\

\hline

\scriptsize{$\bar\nu_{\mu}$ QE(+RPA)+2p2h (Cal) (IH) (Eff=100\%)} & \scriptsize{$0.444$} & \scriptsize{$2.48\times10^{-3}$} & \scriptsize{$1.69252$} & & \scriptsize{$(14)$}\\

\scriptsize{$\bar\nu_{\mu}$ QE(+RPA)+2p2h (Kine) (IH) (Eff=100\%)} & \scriptsize{$0.508$} & \scriptsize{$2.48\times10^{-3}$} & \scriptsize{$0.913228$} & & \scriptsize{$(14)$}\\

\scriptsize{$\bar\nu_{\mu}$ QE (Cal) (IH) (Eff=100\%)} & \scriptsize{$0.46$} & \scriptsize{$2.48\times10^{-3}$} & \scriptsize{$1.46947$} & \scriptsize{$0.566, 2.38\times10^{-3}$(IH)} & \scriptsize{$(15)$}\\

\scriptsize{$\bar\nu_{\mu}$ QE (Kine) (IH) (Eff=100\%)} & \scriptsize{$0.476$} & \scriptsize{$2.48\times10^{-3}$} & \scriptsize{$1.0685$} & & \scriptsize{$(15)$}\\
\hline
\end{tabular}
\end{center}
\caption{Summary of best-fit values and corresponding $\chi^{2}$ values of the oscillation parameters along with global best-fit values for antineutrino for different scenarios studied in this work.}
\label{tab:data3}
\end{table}

The highlights and new findings of this work can be summarised as - using flux and realistic detector specifications of NO$\nu$A in this study, and in all cases that we investigated, the calorimetric method (except for detector efficiency 31.2\% and 33.9\%) produces more precision than the kinematic one, in the parameter sensitivity analysis, for both neutrino and antineutrino and both the mass hierarchies. It is noted that realistic detector effects enhance the uncertainty in the sensitivity analysis in both calorimetric and kinematic methods, while better efficiency of the detector reduces it. Less smearing is observed in migration matrices at ND when the calorimetric method is used (Fig. 3 and 4), than when the kinematic method, which is a clear indication that the former method is more efficient for energy reconstruction. It is observed that at ND, the kinematic method shows more bias in smearing than the calorimetric method when nuclear effects are included (Fig. 3). In sensitivity analysis too, this feature is seen. The dependence of event distribution at ND with randomly chosen BE of target Carbon was shown in Fig. 7, and no significant dependence on BE is observed. When realistic detector specifications are not included, the kinematic method is affected more by the nuclear models (and it depends on the correct understanding of neutrino-nucleus interactions), while the calorimetric method is less biased by them (Fig. 8 and 9). A large contribution to uncertainties due to multi-nucleon enhancement (QE+RPA and 2p2h) both in neutrino and antineutrino cases is observed (Fig. 10 and 11). This contribution is significantly large for the kinematic method than for the calorimetric method, as an ideal detector set-up was used in Figs. 10 and 11. With the inclusion of realistic detector setup details in these two methods (with nuclear effects), it is found that the calorimetric method shows more bias for efficiency 31.2\% and 33.9\% (uncertainty increases, e.g. see Figs. 12-15) in sensitivity contours, as compared to the kinematic method. We did this analysis for both the mass hierarchies as well as for incoming neutrino and antineutrino beams. We note that the quality and quantity of the above biases are almost the same in the two MH for neutrino, and uncertainty is generally less in neutrino than in antineutrino. Through our results in Fig 16, we showed that CP and MH sensitivity have some dependence on energy reconstruction methods, and hence these should be included in such analyses of experimental data carefully.\\

Though our results are in qualitative agreement with those of a similar work \cite{ Ankowski:2015jya}, they are an addition of more detailed and up-to-date results to those of \cite{ Ankowski:2015jya} giving a deeper insight - as we used realistic detector specifications of NO$\nu$A LBL experiment, used a latest complete model for nuclear interactions \cite{Deka:2021qnw} along with long-range RPA corrections and multi-nucleon enhancement, used latest global fit values of known neutrino oscillation parameters, compared the effect of two different values of detector efficiencies, showed migration matrices in details and did the whole analysis for both the mass hierarchies and for both neutrino and antineutrino. These results therefore can be useful for pinpointing neutrino mass hierarchy as well as the measurement of the leptonic CP phase. As the calorimetric method is seen to be less sensitive to nuclear models, thus it appears to be an effective way of minimizing systematic uncertainties in oscillation experiments, though with a careful inclusion of nuclear effects. Of course, one needs to improve the performance of detectors too, for more precise extraction of unknown neutrino oscillation parameters. The results presented in this work can be used as a guiding tool for a careful inclusion of all these effects into neutrino event generators and nuclear models for oscillation parameter sensitivity analysis particularly at NO$\nu$A and in general at all LBL experiments, including leptonic CP phase and MH measurement.

\section{Acknowledgments} 
\label{sec:6}

KB and PD acknowledge support from FIST grant SR/FST/PSI-213/2016(C) dtd. 24/10/2017(Govt. of India) in upgrading the computer laboratory of the Physics department of GU where part of this work was done. KB thanks DST-SERB, Govt. of India for a research project EMR/2014/000296 (2015-2019) during which some initial part of this work was done.

\end{document}